\documentclass[journal]{IEEEtran}
\usepackage[utf8]{inputenc}  
\usepackage{authblk}
\usepackage{graphicx}
\usepackage{xcolor}
\usepackage{cite}
\usepackage{booktabs}
\usepackage{balance}
\usepackage{url}
\usepackage{booktabs}
\usepackage{graphicx}
\usepackage{array}
\usepackage{multirow}
\usepackage{amsmath}
\usepackage[table]{xcolor}
\usepackage{array}
\usepackage{longtable}
\usepackage{supertabular} 

\title{Graph Neural Networks for Vehicular Social Networks: Trends, Challenges, and Opportunities}

\author{
\IEEEauthorblockN{Elham Binshaflout, \textit{Student Member, IEEE}, Aymen Hamrouni, \textit{Student Member, IEEE}, and Hakim Ghazzai, \textit{Senior Member, IEEE}}\\
{\thanks {\hrule
\vspace{0.1cm} 
A part of this work has been published in IEEE International Conference on Smart Mobility (SM'23), Thuwal, Saudi Arabia, Mar. 2023~\cite{Binshaflout_GNN_TPR_2023}. \newline
Elham Binshaflout and Hakim Ghazzai are with the Computer, Electrical and Mathematical Sciences \& Engineering (CEMSE) Division at King Abdullah University of Science and Technology (KAUST), Thuwal, Saudi Arabia (E\textendash mails: \{elham.binshaflout,hakim.ghazzai\}@kaust.edu.sa).\newline
Aymen Hamrouni is with the "Waves: Core Research and Engineering (WaveCore)" group at Katholieke Universiteit Leuven, Belgium (E\textendash mail: \{aymen.hamrouni\}@kuleuven.be).\newline
Elham Binshaflout is also with the College Of Computer Science and Information Technology, Imam Abdulrahman Bin Faisal University, Alkhobar, Saudi Arabia.\newline
}}\vspace{-0.8cm}}

\date{\today}

\graphicspath{{figures/}}

\begin{document}
\maketitle

\begin{abstract}
\textcolor{black}{Graph Neural Networks (GNNs) have emerged as powerful tools for modeling complex, interconnected data, making them particularly well suited for a wide range of Intelligent Transportation System (ITS) applications. This survey presents the first comprehensive review dedicated specifically to the use of GNNs within Vehicular Social Networks (VSNs). By leveraging both Euclidean and non-Euclidean transportation-related data, including traffic patterns, road users, and weather conditions, GNNs offer promising solutions for analyzing and enhancing VSN applications. The survey systematically categorizes and analyzes existing studies according to major VSN-related tasks, including traffic flow and trajectory prediction, traffic forecasting, signal control, driving assistance, routing problem, and connectivity management. It further provides quantitative insights and synthesizes key takeaways derived from the literature review. Additionally, the survey examines the available datasets and outlines open research directions needed to advance GNN-based VSN applications. The findings indicate that, although GNNs demonstrate strong potential for improving the accuracy, robustness, and real-time performances of on task-specific or sub-VSN graphs, there remains a notable absence of studies that model a complete, standalone VSN encompassing all functional components. With the increasing availability of data and continued progress in graph learning, GNNs are expected to play a central role in enabling future large-scale and fully integrated VSN applications.}

\end{abstract}

\begin{IEEEkeywords}
Graph Neural Networks, Vehicular Social Networks, Intelligent Transportation Systems, Smart mobility
\vspace{-0.3cm}
\end{IEEEkeywords}
\section{Introduction}
\IEEEPARstart{D}eveloping an efficient Intelligent Transportation System (ITS) is becoming increasingly demanding due to several factors including, but not limited to, growing urbanization, increased vehicle ownership, and environmental concerns. The rapid increase in traffic volume and industrial city development has also made transitioning from traditional to smart cities increasingly urgent. This involves establishing integrated communication channels among road users and enabling simultaneous processing of tasks, including transportation management and control, seamless vehicle communication, and safety and privacy enhancement~\cite{Mahrez_SmartUrban_2022}.

{\color{black} Within this framework, Vehicular Social Networks (VSNs) emerge as a specialized subset, distinctly enabling real-time communication between various road entities such as vehicles, road users, riders, and road units. VSNs integrate Social Networks (SNs) and Vehicular Ad-hoc Networks (VANETs) to establish a highly efficient platform for data dissemination. As a cutting-edge advancement within the Internet of Vehicles (IoV), VSNs provide a unique platform for reliable communication and collaborative decision-making among vehicles and road participants. This technology aims to enhance driving safety, efficiency, and comfort by enabling vehicles to share information about traffic conditions, road hazards, and driving behaviors in real time~\cite{vegni_survey_2015}. VSNs leverage these interactive capabilities to alleviate traffic congestion, reduce emissions, and improve the responsiveness of drivers, thus offering targeted solutions that complement the overarching goals of ITS\cite{zhang_graph_2022}. The role of VSNs in driving forward these intelligent capabilities highlights their potential to shape the future of vehicular communication significantly.}

{\color{black}Moreover, the integration of VSNs within ITS is further justified by their capacity to handle and analyze the high volume of data generated through these interactive lines of communication. With the increasing availability of high-dimensional real-time traffic data, deep-learning-based approaches, particularly those utilizing graph neural networks, have been proposed to enhance VSN-related tasks~\cite{10038641, essien2021deep, 9756574}. These models are crucial for capturing and analyzing historical traffic data flows, understanding the relationships among traffic participants, and accurately representing the dynamic spatial-temporal metrics of traffic networks. Hence, there is a challenge to accurately represent the traffic features, road network structure, and road users' relationships while extracting the topological relations and complex correlations among the connected entities.}

\begin{table*}[]
\centering
\caption{Existing surveys on GNNs for ITS and related fields and comparisons with the proposed work}
\label{table1listofsurveys} 
\resizebox{1\textwidth}{!}{
\begin{tabular}{|c|c|c|c|c|}
\hline
\textbf{Title} & \textbf{Ref.} & \textbf{Year} & \textbf{Main Focus} & \textbf{Limitations} \\ 
\hline

\begin{tabular}[c]{@{}c@{}}A Survey on Vehicular \\ Social Networks\end{tabular} &
\cite{vegni_survey_2015} & 2015 &
{\color{black}\begin{tabular}[c]{@{}c@{}}Explores established trends in integrating connected \\ vehicles with social networks.\end{tabular}} &
{\color{black}\begin{tabular}[c]{@{}c@{}}Absence of graph learning or GNN-based\\ modeling for VSN applications.\end{tabular}} \\ 
\hline

\begin{tabular}[c]{@{}c@{}}A Survey on Platoon-Based \\ Vehicular Cyber-Physical Systems\end{tabular} &
\cite{jia_survey_2016} & 2016 &
{\color{black}\begin{tabular}[c]{@{}c@{}}Reviews platoon-based vehicular CPS, including \\ techniques and communication standards.\end{tabular}} &
{\color{black}\begin{tabular}[c]{@{}c@{}}Limited focus on VSNs; lack of social interactions,\\ graph-based modeling, or GNN methods.\end{tabular}} \\ 
\hline

\begin{tabular}[c]{@{}c@{}}Vehicular Social Networks: \\ A Survey\end{tabular} &
\cite{rahim_vehicular_2018} & 2018 &
{\color{black}\begin{tabular}[c]{@{}c@{}}Reviews VSN architectures, communication models, \\ and mobility patterns.\end{tabular}} &
{\color{black}\begin{tabular}[c]{@{}c@{}}Lack of graph-learning and absence of \\ GNN-based analysis.\end{tabular}} \\ 
\hline

\begin{tabular}[c]{@{}c@{}}A Survey on Pseudonym Changing \\ Strategies for VANETs\end{tabular} &
\cite{boualouache_survey_2018} & 2018 &
{\color{black}\begin{tabular}[c]{@{}c@{}}Covers security and privacy-preserving strategies \\ in VANETs.\end{tabular}} &
{\color{black}\begin{tabular}[c]{@{}c@{}}Lack of VSN context; absence of graph-learning\\ and GNN-based approaches.\end{tabular}} \\ 
\hline

\begin{tabular}[c]{@{}c@{}}Graph Neural Networks: A Review \\ of Methods and Applications\end{tabular} &
\cite{zhou2021graph} & 2020 &
{\color{black}\begin{tabular}[c]{@{}c@{}}Comprehensive overview of GNN architectures and \\ applications across domains.\end{tabular}} &
{\color{black}\begin{tabular}[c]{@{}c@{}}Does not address ITS or VSN applications;\\ no coverage of transportation datasets.\end{tabular}} \\ 
\hline

\begin{tabular}[c]{@{}c@{}}Network Representation Learning: \\ A Macro and Micro View\end{tabular} &
\cite{liu_network_2021} & 2021 &
{\color{black}\begin{tabular}[c]{@{}c@{}}Analyzes network embedding and graph-based \\ representation learning methods.\end{tabular}} &
{\color{black}\begin{tabular}[c]{@{}c@{}}Absence of ITS/VSN applications; \\ does not explore GNN-based solutions.\end{tabular}} \\ 
\hline

\begin{tabular}[c]{@{}c@{}}Machine Learning for Next-Generation \\ Intelligent Transportation Systems\end{tabular} &
\cite{tingting_machine_2021} & 2021 &
{\color{black}\begin{tabular}[c]{@{}c@{}}Reviews ML techniques in ITS (cooperative driving,\\ hazard detection, etc.).\end{tabular}} &
{\color{black}\begin{tabular}[c]{@{}c@{}}Lack of graph-learning perspective; does not \\ address GNN or VSN-specific modeling.\end{tabular}} \\ 
\hline

\begin{tabular}[c]{@{}c@{}}Graph Neural Networks: \\ Architectures, Stability, and Transferability\end{tabular} &
\cite{ruiz_graph_2021} & 2021 &
{\color{black}\begin{tabular}[c]{@{}c@{}}Focuses on GNN robustness and transferability \\ across graph structures.\end{tabular}} &
{\color{black}\begin{tabular}[c]{@{}c@{}}Does not explore ITS or VSN tasks;\\ no discussion of mobility or social graphs.\end{tabular}} \\ 
\hline

\begin{tabular}[c]{@{}c@{}}A Comprehensive Survey on \\ Graph Neural Networks\end{tabular} &
\cite{wu_comprehensive_2021} & 2021 &
{\color{black}\begin{tabular}[c]{@{}c@{}}Compares GNN architectures, datasets, open-source \\ implementations, and applications.\end{tabular}} &
{\color{black}\begin{tabular}[c]{@{}c@{}}Does not address ITS or VSN domains.\end{tabular}} \\ 
\hline

\begin{tabular}[c]{@{}c@{}}Graph Neural Networks in Recommender \\ Systems: A Survey\end{tabular} &
\cite{wu_graph_2022} & 2022 &
{\color{black}\begin{tabular}[c]{@{}c@{}}Reviews GNN-based recommender systems and their\\ learning paradigms.\end{tabular}} &
{\color{black}\begin{tabular}[c]{@{}c@{}}No coverage of transportation or VSN applications.\end{tabular}} \\ 
\hline

\begin{tabular}[c]{@{}c@{}}Graph Neural Network for Traffic \\ Forecasting: A Survey\end{tabular} &
\cite{jiang_graph_2022} & 2022 &
{\color{black}\begin{tabular}[c]{@{}c@{}}Summarizes spatio-temporal GNN models \\ for traffic forecasting.\end{tabular}} &
{\color{black}\begin{tabular}[c]{@{}c@{}}Limited to forecasting; lack of VSN context;\\ absence of social-interaction graph modeling.\end{tabular}} \\ 
\hline

\begin{tabular}[c]{@{}c@{}}Data-Centric Approaches in the \\ Internet of Vehicles\end{tabular} &
\cite{partovi2023data} & 2023 &
{\color{black}\begin{tabular}[c]{@{}c@{}}Reviews data management and analysis in IoV.\end{tabular}} &
{\color{black}\begin{tabular}[c]{@{}c@{}}Focus on data management; lack of GNN \\ or social-vehicular graph modeling.\end{tabular}} \\ 
\hline

\begin{tabular}[c]{@{}c@{}}The Internet of Vehicles and Sustainability:\\ Reflection on Environmental, Social, \\ and Corporate Governance\end{tabular} &
 \cite{kostrzewski2023internet} &
  2023 &
  {\color{black}\begin{tabular}[c]{@{}c@{}}Proposing a sustainable model for IoV design,\\ mapping Environmental, Socia,l and Governance factors \end{tabular}} &
  {\color{black}\begin{tabular}[c]{@{}c@{}} Absence of graph learning or GNN modeling \end{tabular}} \\ 
\hline

\begin{tabular}[c]{@{}c@{}}Graph Neural Networks for Traffic \\ Pattern Recognition\end{tabular} &
\cite{Binshaflout_GNN_TPR_2023} & 2023 &
{\color{black}\begin{tabular}[c]{@{}c@{}}Overview of GNNs for traffic pattern analysis.\end{tabular}} &
{\color{black}\begin{tabular}[c]{@{}c@{}}Limited to one domain; lack of VSN-level\\ graph modeling or multi-domain coverage.\end{tabular}} \\ 
\hline

\begin{tabular}[c]{@{}c@{}}Graph Neural Networks for Time Series\end{tabular} &
\cite{jin2024survey} & 2024 &
{\color{black}\begin{tabular}[c]{@{}c@{}}Analyzes GNN architectures for time-series \\ forecasting, classification, and anomaly detection.\end{tabular}} &
{\color{black}\begin{tabular}[c]{@{}c@{}}Does not address ITS or VSN tasks; absence \\ of social or mobility graph modeling.\end{tabular}} \\ 
\hline

\begin{tabular}[c]{@{}c@{}}Graph Neural Networks in Intelligent \\ Transportation Systems\end{tabular} &
\cite{li2024graphneuralnetworksintelligent} & 2024 &
{\color{black}\begin{tabular}[c]{@{}c@{}}Reviews GNN applications across six ITS domains.\end{tabular}} &
{\color{black}\begin{tabular}[c]{@{}c@{}}Lack of focus on VSN; limited social-interaction\\ graph modeling; datasets not categorized\\ by VSN-specific tasks.\end{tabular}} \\ 
\hline

\begin{tabular}[c]{@{}c@{}}Graph Neural Networks for \\ Vehicular Social Networks: \\  Trends, Challenges and Opportunities\end{tabular} &
  \textbf{Proposed} &
  \textbf{$-$}&
  \textbf{{\color{black}\begin{tabular}[c]{@{}c@{}}Comprehensive review and analysis \\of VSNs and the GNNs architectures.\\ Present an overview of promising GNN-based VSN\\ applications, structures, and datasets.\end{tabular}}} &
  \textbf{{\color{black}\begin{tabular}[c]{@{}c@{}}Bridges the gap between VSN and GNN research by\\ jointly discussing both domains and providing a\\ comprehensive review of GNN-based VSN applications,\\ categorized datasets, and key future opportunities.\end{tabular}}} \\ 
\hline

\end{tabular}}
\end{table*}

With the high volume and heterogeneity of the data, this task becomes non-trivial to handle by vanilla deep learning technologies. As a matter of fact, such traditional technologies usually face the challenge of increasing complexity when applied to real-time data, which affects the accuracy and training time. It has been shown that those methods lack the ability to deal with uncertainty in the data captured, and this is one of the main features of traffic data structure~\cite{Tijs_Uncertainty_2020}. Graph Neural Networks (GNNs) are novel architectures in deep learning technologies that have been widely applied in a variety of areas, including recommendation systems, service discovery in Internet-of-Things (IoT) networks, and bioinformatics~\cite{9446513,hamrouni_low-complexity_2022,9859333}. This novel paradigm can represent and manipulate graph-based data, i.e., non-Euclidean data, more effectively. Recently, they have become a very popular tool for modeling and recognizing traffic patterns. GNNs can model complex relationships between the different traffic elements, such as intersections, roads, and vehicles, allowing for effective representation and analysis of large-scale traffic data. There are various GNN-based approaches that have been proposed and applied to a number of traffic-related tasks, including traffic flow prediction~\cite{chen_aargnn_2022}, traffic anomaly detection~\cite{Wu_GNN_Anomaly_2022}, and traffic control optimization~\cite{Zhong_probablistic_GNN_2021}. In~\cite{Binshaflout_GNN_TPR_2023}, we have provided a brief review of the use of GNNs for traffic pattern recognition applications. All the previous studies have demonstrated the effectiveness of GNNs in handling the complexity and heterogeneity of real-world traffic data, leading to improved performance compared to traditional methods. 

This survey aims to present a comprehensive review of the applications of GNN on VSN. It provides a broader focus on VSN-related applications to improve ITS and smart mobility services. In this survey, we shed light on the most widely utilized GNN architectures in transportation research. We then explore the various techniques employed in the literature for diverse VSN-related applications, including traffic pattern recognition, flow and trajectory prediction, driving assistance, and traffic data analysis. It also enumerates the datasets existing in the literature that can be used to model VSN-related applications. Finally, the survey examines the potential avenues to further advance the field. It is worth mentioning that there are a limited number of existing reviews on the topic of GNNs~\cite{wu_comprehensive_2021} as well as on VSN in general. \textcolor{black}{The main contributions of this survey are summarized as follows:}
\begin{itemize}
    \item \textcolor{black}{We present the first comprehensive survey dedicated to the use of GNNs within VSNs, highlighting how graph-based learning supports and enhances a wide range of VSN-related applications.}
    \item \textcolor{black}{We propose a structured categorization of existing studies across core VSN tasks, including traffic flow and trajectory prediction, traffic forecasting, signal control, routing, driving assistance, as well as connectivity management, and review the GNN architectures used to address them.}
    \item \textcolor{black}{We perform a quantitative analysis of the surveyed studies and identify key takeaways about the usage of GNNs for VSN-related applications. We also conduct a detailed analysis of open-source datasets, methodological choices, and trends in model performance.}
    \item \textcolor{black}{We outline the current research gaps and synthesize emerging challenges and future directions needed to enable scalable, real-time, and socially aware VSN systems.}
\end{itemize}

\textcolor{black}{As presented in Table~\ref{table1listofsurveys}, to the best of our knowledge, there is no existing review covering the topic of GNNs and their existing applications in VSN. The majority of them concentrate on these subjects independently. For example, the authors of~\cite{tingting_machine_2021} reviewed the existing machine-learning approaches used to solve a range of general ITS applications. The authors of~\cite{jiang_graph_2022} provided a survey on different GNN architectures that focuses exclusively on traffic forecasting problems. In~\cite{wenjuan_research_2022}, the authors focused on GNN-based traffic flow prediction models. Hence, unlike broader surveys that encompass a wide spectrum of ITS applications, this survey specifically targets the use of GNNs within VSNs, offering a detailed exploration of how these networks exploit complex graph structures derived from vehicle connectivity and social interactions. This focus allows us to delve deeply into the unique challenges and opportunities presented by VSN-related applications, distinctly setting our work apart from general ITS surveys, such as~\cite{li2024graphneuralnetworksintelligent}, by emphasizing specialized applications rather than a generalized overview.}

\begin{figure*}[tp]
    \centering
    \includegraphics[width=17cm]{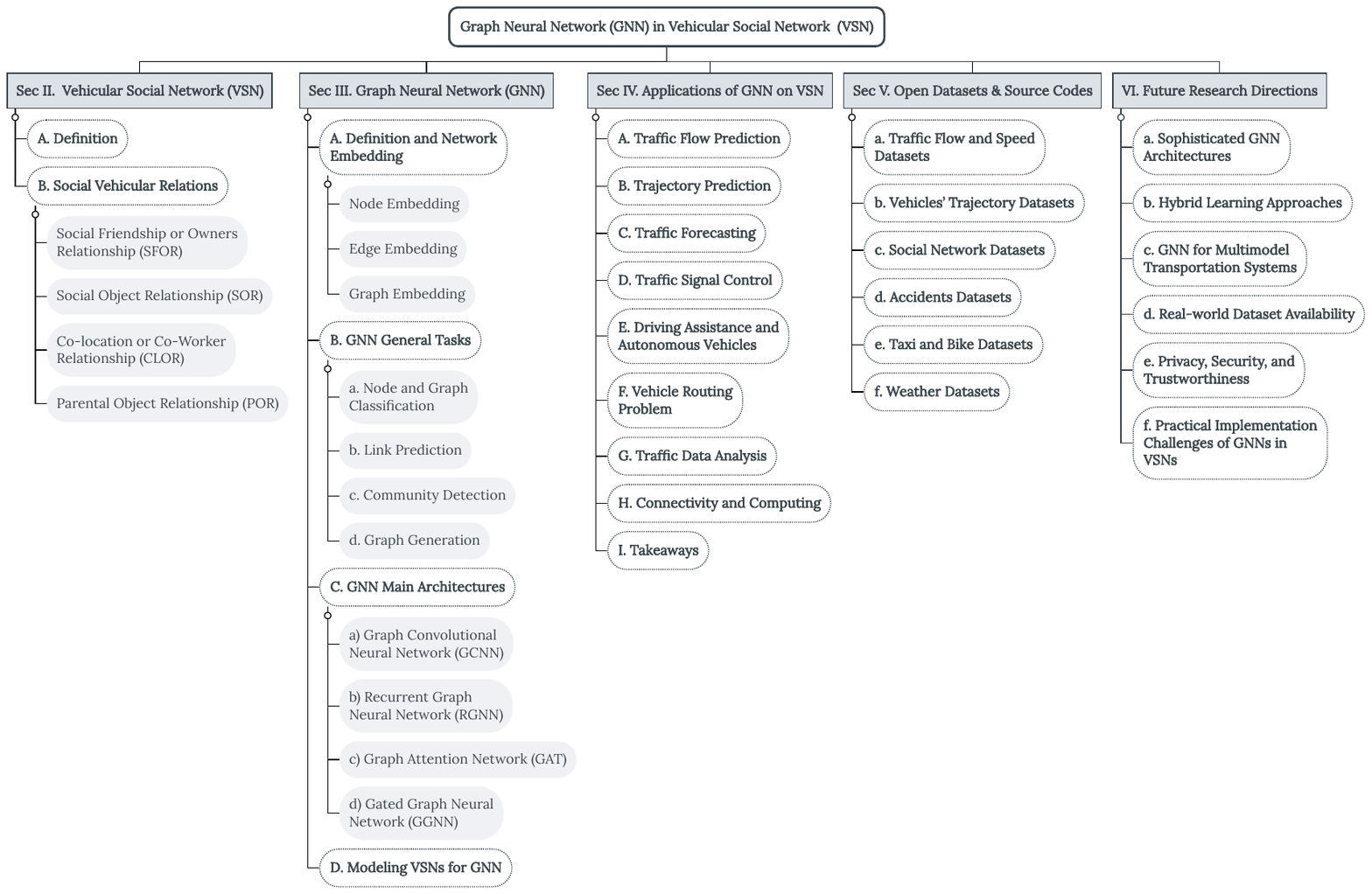}
  \caption[width=14cm]{\color{black}The structure of this survey paper.}
        \label{fig:GNN_VSN paper}
\end{figure*}

The rest of the paper is organized as follows: Section~\ref{sec2} introduces the reader to the concept of VSN, its potential, and its main challenges. Section~\ref{sec3} introduces the GNNs\footnote{\textcolor{black}{Readers familiar with the foundational concepts of VSN and GNNs may choose to skip Section II and Section III and proceed directly to Section IV.}}, their general learning tasks, and their architectures that are relevant to the ITS/VSN problems. Section~\ref{sec4} addresses the applications of GNN on VSN with examples of the current advancement in the field and GNN models that are applied under each of the VSN tasks. Section~\ref{sec5} provides a list of the currently existing open-source datasets that are available for experiments. Section~\ref{sec6} provides possible future directions in the area, and finally, Section~\ref{sec7} concludes the survey. Fig.~\ref{fig:GNN_VSN paper} provides a high-level overview of the content of the survey.

\section{Vehicular Social Networks}
\label{sec2}
\subsection{Definition}
VSNs are wireless networking systems that allow communication between different road entities (e.g., vehicles, road users, riders, and road units) in real-time. It integrates Social Networks (SNs) and the Vehicular Ad-hoc NETworks (VANETs) to establish a highly efficient data dissemination~\cite{rahim_vehicular_2018}. They typically utilize ITS technologies and IoT networks to enable short-range communication between nearby vehicles as well as a wide area network to allow longer-distance sharing of useful traffic data. By connecting heterogeneous road user entities such as vehicles, pedestrians, and connected infrastructure in a collaborative system, they can provide accurate and up-to-date information on traffic connections, incident notifications, and the status of other drivers in their vicinity~\cite{ning_vehicular_2017}.

\textcolor{black}{By integrating SNs, VSNs can include human elements that enhance the vehicular connection in addition to the traditional Vehicle-to-Vehicle (V2V) and Vehicle-to-Infrastructure (V2I) communication links~\cite{ning_vehicular_2017}. This integration enhances ITS applications by enabling traffic and road condition sharing, providing real-time hazard alerts, and helping riders find faster routes in cities. Furthermore, the integration of VSN with other technological domains enables a wider range of services such as healthcare, theft prevention, emergency alert and response, road safety, and entertainment (e.g., sharing news, playing games, or reading). In the context of environmental enhancement, VSN can help reduce energy consumption and CO$_2$ emissions by offering better monitoring and cooperation between road users. }

\subsection{Social Vehicular Relations}
\textcolor{black}{Integrating social networking into IoV has led to the Social Internet of Vehicles (SIoV), extending the social IoT concept~\cite{hosseinzadeh2024advancing}. SIoV identifies possible social relations among road objects and users in VSN~\cite{vegni_survey_2015,Khanfor_Automated_2020}, fostering trustworthy collaboration and services. In the following, we include some examples of these relationships:
\begin{itemize}
    \item \textit{Social Friendship or Owners Relationship (SFOR):}
    This relationship can be established when two or more ITS devices (e.g., vehicles, sensors, etc.) share the same owner, or their owners have a social relationship. An example of the first case is vehicles owned by the same person, or owned by a corporation. The second case is when owners are friends or relatives (e.g., they follow each other on social media or have similar interests).
    \item \textit{Social Object Relationship (SOR):}
    This is established when road users communicate in V2V links such as collaborating on routing or information sharing.  SOR relationship strength depends on user trajectories, locations, and communication history.
    \item \textit{Co-location or Co-Worker Relationship (CLOR):}
    This relation is generated among ITS devices that are frequently located in the same geographical area like vehicles on the same highway or passing Road Side Units (RSU). These devices form a CLOR relation to share useful life data about their speed and road status if they are geographically close enough to each other with a distance less than a certain threshold.
    \item \textit{Parental Object Relationship (POR):} This relation can be created between devices that are developed by the same manufacturer, e.g., same brand vehicles having similar features or produced in the same period (same generation). This enables sharing information about the vehicle status and performing diagnostic and remote maintenance services.
\end{itemize}}

\textcolor{black}{Such social relations (among others) can be established between road users and objects to create meaningful VSNs that can be utilized to extract accurate information, consequently improving ITS and smart mobility services. Nonetheless, the concept of VSN still faces challenges in modeling successful algorithms capable of handling the heterogeneous and dynamic structure of its networks. In fact, in such highly dynamic, ubiquitous, and large-scale networks, enabling effective network understanding, service discovery, and navigability of VSN is not straightforward especially, when operating in real-time for many delay-intolerant applications. Since VSNs can be easily modeled using graphs, the pioneering GNNs generally present a promising technology for investigating complex structures and designing relevant solutions. Thanks to their graph structure learning capabilities, GNNs can be very effective in leveraging many ITS services applied to VSN and performing rapid decisions during inference time. In the next section, we provide a comprehensive introduction to the GNN paradigm, as well as the general GNN tasks applicable to VSN. We also provide an overview of the different model architectures that can be utilized in VSN graphs.}
\section{Graph Neural Networks}
\label{sec3}


\textcolor{black}{Graphs are effective for representing complex, unstructured real-world data, especially in non-Euclidean spaces, where traditional neural networks fail to optimize solutions. For example, molecules in drug discovery and user-product relationships in e-commerce are best modeled as graphs for efficient analysis and accurate recommendations. This has sparked growing interest in improving deep learning methods for graph data across various settings, such as supervised, unsupervised, and reinforcement learning~\cite{zhou2021graph}.}

\textcolor{black}{GNNs are a new paradigm that has recently evolved in response to this need. It enables neural networks to function on any graph-based data type. In this context, a considerable number of works investigated various strategies to enhance existing machine learning approaches, such as Convolutional Neural Networks (CNN), Recurrent Neural Networks (RNN), and Autoencoders to eventually produce generalizing models that can also handle graph data. The results of these studies presented a promising future for this technology, especially when working with decentralized data structures~\cite{wu_graph_2022}. In this section, we present a thorough overview of the various standard tasks and architectures of GNN models in their raw forms to provide researchers and practitioners with an unbiased introduction to the GNN concept. In the next section (Section~\ref{sec4}), we will investigate how GNNs can be applied to VSNs or part of them to leverage ITS applications.}

\subsection{Definition and Network Embedding}

A graph consists of a set of nodes (or entities) and edges (or relationships between nodes). It can be denoted as $G = (V,E)$, where $V$ is the set of vertices and $E$ is the set of edges in the graph. The graph can represent different types and scales (e.g., directed or undirected, homogeneous or heterogeneous, and static or dynamic, etc) depending on the context of the problem and structure of the data~\cite{zhou2021graph,liu_network_2021}. The GNN is a type of neural network that operates on these graph-structured data. It takes the graph as input and performs operations on the nodes' and edges' features to capture the complex relationships between the nodes and network topology.

\begin{figure}[t]
        \centering
        \includegraphics[width=8.5cm]{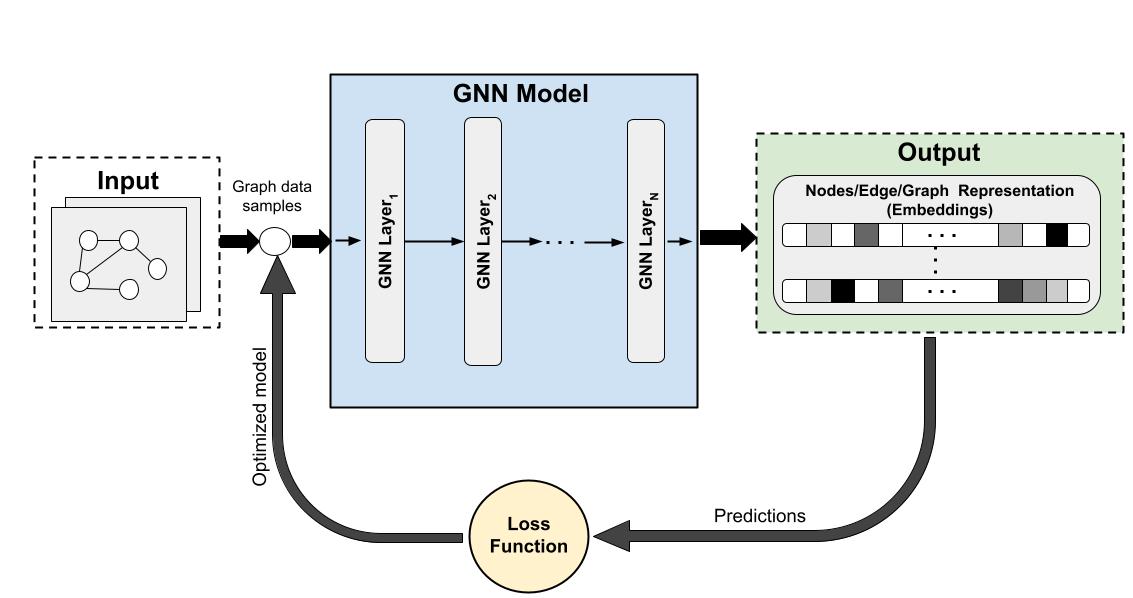}
        \caption{Vanilla architecture of a GNN showing the basic steps of transforming the input graph into embeddings.}
        \label{fig:GNN general pipeline}\vspace{-0.4cm}
\end{figure}

\textcolor{black}{GNNs are essentially employed to generate numerical representation from graph-structured data. They convert a graph or a network of graphs into a lower dimensional continuous latent space (i.e., a numerical representation) while extracting relevant information from the graph and learning how nodes, edges, and structures are correlated through propagation.  The goal of this transformation is to create an output in the vector spaces, as the latter is more amenable to data science tools than complex graphs formed by edges and vertices. This complexity reduction of the data structure and the transformation of the data representation into matrices indeed accelerates the computation time and increases the efficiency of the model. In Fig.~\ref{fig:GNN general pipeline}, we present a generalized architecture of the input/output process of the GNN model. }

\begin{figure}[tp]
    \centering
    \includegraphics[width=9cm]{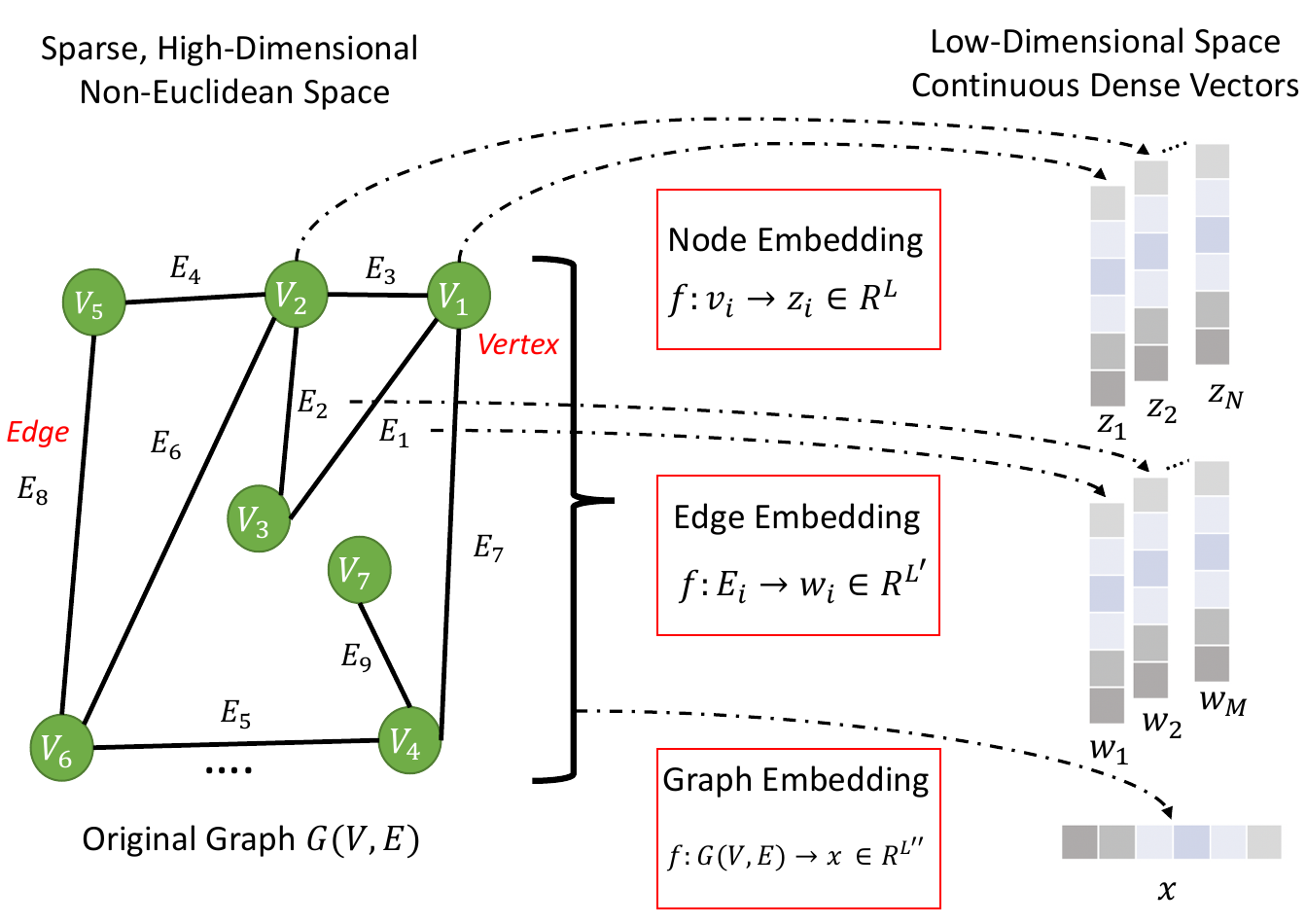}
    \caption{Illustration of the different types of embeddings: Node Embedding, Edge Embedding, and Graph Embedding. Each embedding results in a low-dimensional space representation.}
    \label{fig:embeddingtypes}\vspace{-.5cm}
\end{figure}

\textcolor{black}{Regardless of the task domain, the input graph, which can be structured (e.g., social networks, knowledge maps, etc) or unstructured (e.g. images, text, etc), passes a series of neural network layers. In the case of an unstructured scenario, an extra step has to be done, usually prior to the execution of the model, which is to build the graph structure from the task (e.g., converting an image to a graph, etc). Each layer of the GNN model performs a set of computational modules (i.e., sampling, CNN/RNN-based operation, pooling, and skip connection). The deep learning modules are selected based on the design of the GNN architecture, which will be covered in a subsequent subsection. Generally, the process of converting the graph into lower dimension space is known as \textit{network embedding}. Depending on the target of the embedding task, there are three distinguished levels:  1) \textit{node embeddings}, 2) \textit{edge embeddings}, and 3) \textit{graph embeddings}. Fig.~\ref{fig:embeddingtypes} illustrates the embedding process converting a high-dimensional non-euclidean space into a lower-dimensional continuous vector. In the following, we explain each type of embedding:}  




%

\textcolor{black}{\begin{itemize}
    \item \textit{Node Embedding:} Each vertex (or node) is encoded with its vector representation. This approach is used if the objective is to perform visualization or prediction on the vertex level, e.g., visualization of vertices in the 2D plane or prediction of new connections based on vertex similarities. The nodes close to each other in the graph must be close to each other in the embedding space. 
    Such embedding techniques are used to solve problems where the target is to perform predictions, visualization, or classification based on the distribution of the nodes within the network, their relationships, and their characteristics~\cite{hamrouni_low-complexity_2022}. An example of node embedding frameworks are GraphSage~\cite{hamilton2018representation}, DeepWalk~\cite{Perozzi_DeepWalk_2014}, and LINE~\cite{Tang_LINE_2015}. This type of embedding is illustrated in Fig.~\ref{fig:embeddingtypes}, where each node $V_i$ in an $N$-node graph is encoded into a low-dimensional space vector in $R^L$, with $L$ being the size of the embedding space.
    \item \textit{Edge Embedding:} The aim is to embed the edges, not the nodes, by mapping their similarity from the original graph or network of graphs. In fact, the edges are transformed into lower-dimensional vectors by focusing on local and global structure information of the graphs' edges. Instead of finding a mapping of nodes with similar contexts (i.e., node embedding), the objective is to map edges into the embedding space such that the edges that share the same nodes are close. The edge-level embedding has a number of applications, such as social network analysis and biomedical knowledge discovery. As an example, we can cite the Edge2Vec~\cite{Wang_edge2vec_2020}. This type of embedding is illustrated in Fig.~\ref{fig:embeddingtypes}, where each edge $E_i$ in an $M$-edge graph is encoded into a low-dimensional space vector in $R^{L^{'}}$, with $L^{'}$ being the size of the embedding space.
    \item \textit{Graph Embedding:} The entire graph is embedded with a single representation, including the nodes and edges. The learning model captures the overall structure of the graph by learning the node distribution, the relationship between the nodes, and the characteristics of the nodes. Indeed, graph embedding combines edge-level embedding and node-level embedding to derive a single vector representation for the entire graph, which is capable of capturing the interrelated data~\cite{hamrouni_low-complexity_2022}.  This type of embedding is illustrated in Fig.~\ref{fig:embeddingtypes}, where the entire graph $G$ is encoded into a single low-dimensional space vector in $R^{L^{''}}$, with $L^{''}$ being the size of the embedding space.  The applications of this are wide and include analyzing any graph-based problem to perform visualization, graph network analysis, clustering, or classification tasks based on analyzing the whole graphs, sub-graphs, or networks of graphs. Examples of graph embedding frameworks are Graph2vec~\cite{Annamalai_graph2vec_2017}, Walklets~\cite{Bryan_walklets_2016}, and GEM~\cite{Palash_GEM_2018}. 
\end{itemize}}

\begin{figure}[t]
    \centering
    \includegraphics[width=7.5cm]{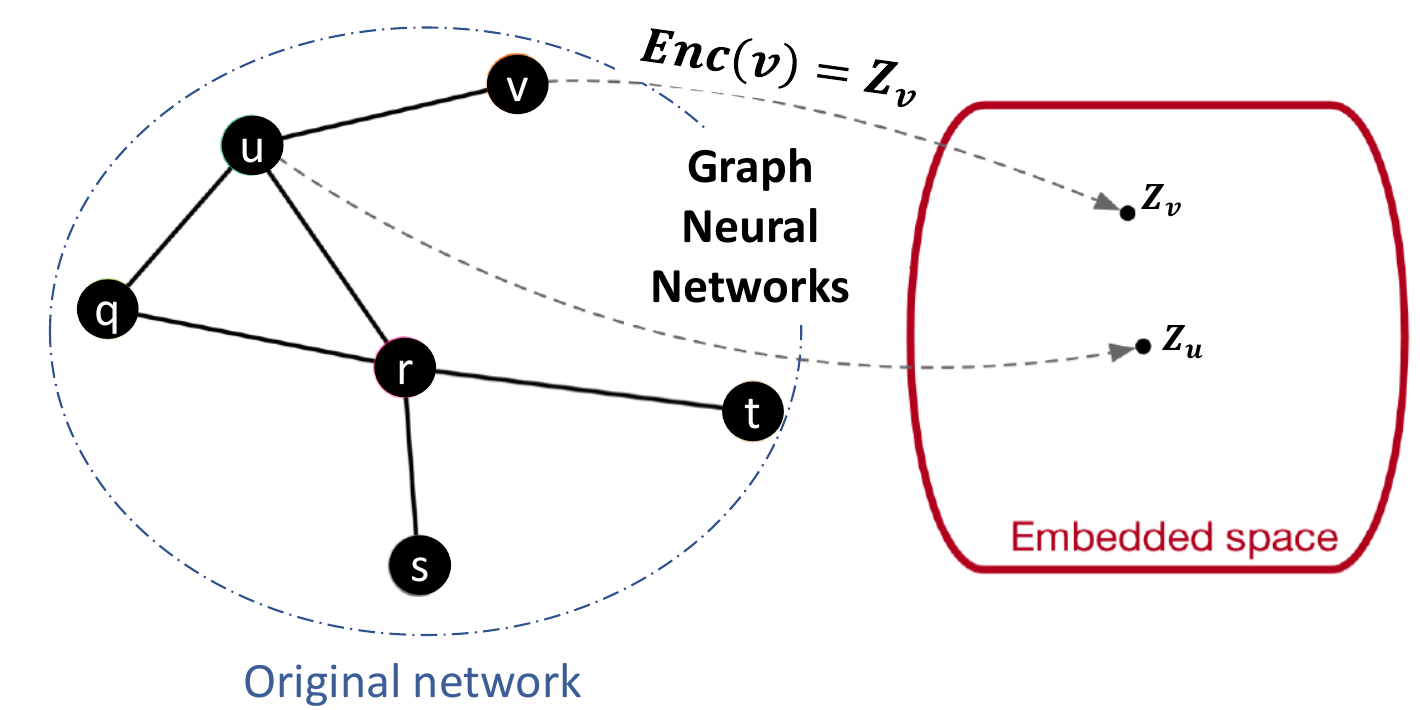}
    \caption{Visualization of the embedding process where the original 3-D network is transformed into an embedding space with 2-D vectors by the means of an encoding function $ENC()$.}
   \label{fig:embeddingexplain}
\end{figure}

\textcolor{black}{The key distinction between these approaches lies in the encoding function, which defines how the graph is mapped into the embedding space. As Fig.~\ref{fig:embeddingexplain} illustrates, this function, referred to as $ENC()$, is responsible for outputting the node $u$'s embedding vector value $Z_u$ from the original network. The resultant vector value $Z_u$  can include either the vertex-to-vertex relationship or both the vertex-to-vertex relationship and the nodes' features. An example of such a mapping function can be a simple encoder that captures only the graph's relationships~\cite{hamilton2018representation}:
\begin{align}
    ENC(u)= \mathbf{z}_{u} = Z \times e_u,
\end{align}
where $Z$ is the output matrix with dimension $d \times |\mathcal V|$ containing the embedding values of the nodes $V$ and $e_u$ is an indicator vector that has all values set to zero except in one column, set to one, indicating the presence state of node $u$.}

To incorporate both node relationships and attributes, the encoder is implemented as an unsupervised multi-layer neural network~\cite{zhou2021graph}. As shown in Fig.~\ref{fig:gnnillustrations}, the first layer takes the input vector $X_{u,0}$ of node $u$, which includes its attributes. The encoding is computed using $X_{u,0}$ and the features of its neighbors $N(u)$, e.g., nodes $q$, $r$, and $v$ in Fig.~\ref{fig:embeddingexplain}. The final layer (e.g., Layer 2) outputs the node embedding. Each layer’s output is denoted $h_u^k$, recursively defined over $k$ layers of aggregation. The neural layers corresponding to layer $k$ for a node $u$ are defined as $h_{u}^{k}$ and can be written as follows:

\begin{figure}[t]
    \centering
    \includegraphics[width=8.75cm]{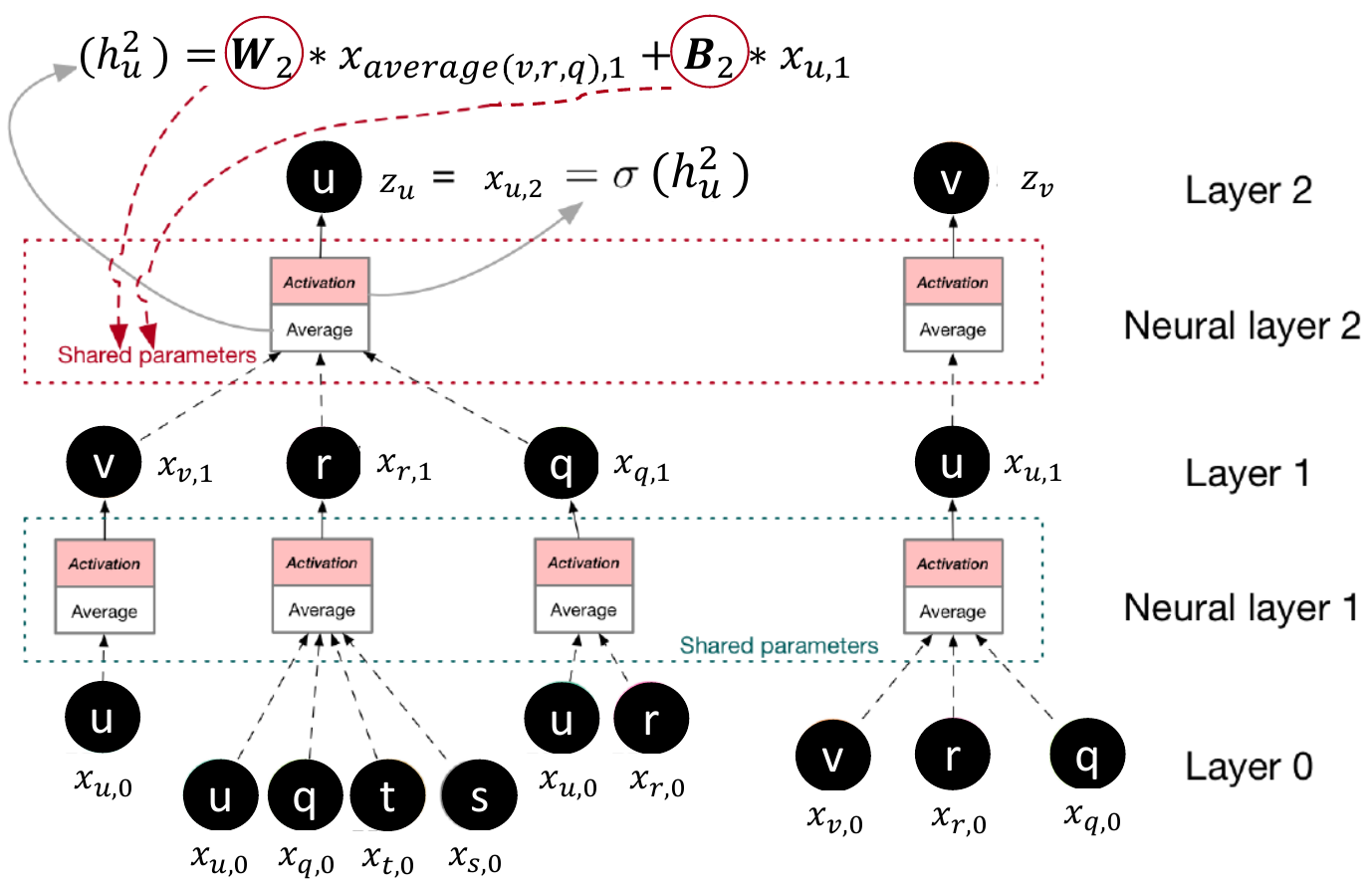}
    \caption{Illustration of the graph neural network embedding two nodes $u$ and $v$ using their attribute vector $x_{u,0}$ and $x_{v,0}$. }
\label{fig:gnnillustrations}
\end{figure}

\begin{align}
&\mathbf{h}_{u}^{k}=&\sigma\left(\mathbf{W}_{k} \sum_{a \in N(u)} \frac{\mathbf{h}_{a}^{k-1}}{|N(u)|}+\mathbf{B}_{k}\notag \mathbf{h}_{u}^{k-1}\right), \\&& \forall k \in\{1, \cdots, K\} 
\end{align}
The initial 0-th layer embedding $\mathbf{h}_{u}^{0}$ are equal to the node features $\mathbf{x}_{u}$ and the optimized embedding $\mathbf{z}_{u}$ are equal to the final layer embedding $\mathbf{h}_{u}^{K}$. The $\sigma$ function represents the non-linearity (e.g., relu), and  $\mathbf{W}_{k}$ and $\mathbf{B}_{k}$ represent the trainable weight matrices that will be adjusted with the loss function. The term $\sum_{a \in N(u)} \frac{\mathbf{h}_{a}^{k-1}}{|N(u)|}$ represents the average of neighbors's previous layer embedding. A more simplified vector format is written as follows:
\begin{equation}
\begin{array}{l}
\mathbf{H}^{(l+1)}=\sigma\left(\mathbf{H}^{(l)} \mathbf{W}^{(l)}+\tilde{\mathbf{A}} \mathbf{H}^{(l)} \mathbf{B}^{(l)}\right),
\end{array}
\end{equation}
where $\tilde{\mathbf{A}}=\mathbf{D}^{-\frac{1}{2}} \mathbf{A} \mathbf{D}^{-\frac{1}{2}}$ and $\mathbf{H}^{(k)}=[\mathbf{h}_{1}^{(k)^{T}}, \ldots,\mathbf{h}_{m}^{(k)^{T}}]^{T}$, where $\mathbf{D}$ is the  diagonal matrix, $\mathbf{A}$ is the social relationship adjacency matrix and $m$ represents the number of embeddings in layer $k$.

After defining the appropriate encoders for each type of embedding, appropriate similarity functions~\cite{zhou2021graph} must also be defined in order to specify how the nodes characteristics in the vector space map to the nodes' characteristics in the original network. 
The similarity between two nodes $u$ and $v$ in the embedding space is simple and can be defined as $\mathbf{z}_{v}^{\top} \mathbf{z}_{u}$, the dot product between the vectors. The goal is to find a function $S$ where the embedding $z_u$ and $z_v$ of the nodes $u$ and $v$ can be optimized in a way such that:
\begin{equation}
S(u, v) \approx \mathbf{z}_{v}^{\top} \mathbf{z}_{u}.
\label{simembedding}
\end{equation}

In the literature, there have been some studies~\cite{Wu_2021,shao2021learning} focusing on multiple similarity functions in the original network space. Some functions include a basic adjacency matrix, a multi-hop network similarity matrix, or a random walk similarity~\cite{9039675,9339909,9306765}. In the random walk technique, the node's similarity in the graph is computed based on the probability that two nodes $u$ and $v$ co-occur on a random walk over the network. Moreover, the probability of visiting node $v$ on a random walk starting from node $u$ using a random walk strategy $R$ describes the similarity between the two nodes. This type of similarity function incorporates both local and higher-order neighborhood information and does not need to consider all node pairs when training; it only needs to consider pairs that co-occur on random walks.
The optimized similarity, or loss, in this case, is written as follows:
\begin{equation}
\mathcal{L}=\sum_{u \in V} \sum_{v \in N_{R}(u)}-\log \left(P\left(v \mid \mathbf{z}_{u}\right)\right),
\end{equation}
where $N_R(u)$ describes the multiset of nodes visited on random walks starting from node $u$ and  $P\left(v \mid \mathbf{z}_{u}\right)$ represent the likelihood of random walk co-occurrences between the node $v$ and embedding of node $u$ computed using Softmax as follows:
\begin{equation}
P\left(v \mid \mathbf{z}_{u}\right)=\frac{\exp \left(\mathbf{z}_{u}^{\top} \mathbf{z}_{v}\right)}{\sum_{n \in V} \exp \left(\mathbf{z}_{u}^{\top} \mathbf{z}_{n}\right)}.
\end{equation}
This term represents the predicted probability of $u$ and $v$ co-occurring on a random walk. Optimizing random walk embeddings means finding  $z_u$  that minimizes $\mathcal L$.

\subsection{\textcolor{black}{GNN General Tasks}}
Embedding, or representation learning, is foundational for GNN tasks. It accelerates graph generation by embedding target outputs early. Traditional methods like matrix factorization and shallow neural embedding struggle with large, dynamic graphs~\cite{liu_network_2021}, whereas GNN-based models excel in handling complex, evolving graphs across fields like physics, chemistry, and social networks~\cite{guo_mixed_2022, hamrouni_low-complexity_2022}. This section explores GNN tasks, their goals, processes, and applications, particularly in ITS.

\paragraph{Node and Graph Classification} 
This task focuses on predicting labels for new, unlabeled nodes or graph instances. Inputs include node features, graph instances, and the structural information of neighboring nodes. GNNs generate node embeddings through iterative message passing, which are then used by a classifier to predict labels. This process leverages trainable parameters optimized via backpropagation~\cite{hamilton2018representation}. Node classification typically follows a semi-supervised approach, requiring only a few labeled nodes to infer labels for new ones, as seen in social network analysis. Compared to traditional neural networks, GNNs significantly improve classification performance, especially in dynamic graph settings~\cite{ma_streaming_2020}.

Graph classification, in contrast, predicts a label for an entire graph. Here, inputs are individual graphs within a larger network of graphs. This hierarchical setup is common in applications like biological networks, document and text classification, and drug discovery. A key challenge is handling the complexity of graph structures. Graph embedding helps address this by converting graphs into vectors that feed into supervised or semi-supervised classifiers for label prediction~\cite{li_semi-supervised_2019}.

\paragraph{Link Prediction} 
This task determines the relationship between two nodes in a graph with an incomplete adjacency matrix. By learning node features, the model forecasts links for new nodes based on historical data~\cite{hamilton2018representation}. 
Link prediction is one of the most utilized GNN tasks for industrial applications. For example, in biology, the structure of the protein or drug interaction can form an automatically expanding graph with many unconnected nodes. Likewise, in e-commerce systems, companies rely on the links between nodes in the graphs to build accurate recommendation systems to suggest new products or movies that match the user interests~\cite{houyoutube,cen_representation_2019}. In social network analysis, links represent the relations between users and their friends and interests in the networks to further recommend friends, posts, ads, etc.~\cite{wu_graph_2022}.
\paragraph{Community Detection} 

Community detection clusters nodes based on edge structures and graph topology. Despite extensive research, it remains a complex and partially unsolved deep learning task. GNNs offer promising solutions across various domains. For instance, in cyber-criminal networks, GNNs can learn hacker behavior to automate security incident detection. In social networks, models like STGSN~\cite{min_stgsn_2021} use spatio-temporal data and attention mechanisms to track evolving communities. Other methods, such as low-complexity recruitment algorithms, combine graph embeddings and clustering for applications like mobile crowdsourcing~\cite{hamrouni_low-complexity_2022}.

\paragraph{Graph Generation} Graph generation aims to create new graphs that resemble a given set by learning the underlying graph distribution. GNNs are well-suited for this, as they can capture implicit patterns in node relationships and structures. Several generative models, such as NetGAN, GraphRNN, and GraphAF, have been introduced for this task~\cite{zhou2021graph}. These models learn from input graph samples, including node and edge features, and generate similar graphs for future data. Applications include recommendation systems~\cite{wu_graph_2022} and molecular graph generation~\cite{shi_graphaf_2020}, where the model predicts molecule structures based on training graphs.

GNNs are highly versatile and can perform multiple tasks with the same architecture. For example, DyGNN supports both link prediction and node classification in dynamic graphs like social networks and VSNs~\cite{ma_streaming_2020}. In complex domains such as molecular analysis, multiple tasks—node classification, graph classification, and graph generation—are often needed simultaneously~\cite{wu_comprehensive_2021}. This highlights the importance of understanding the capabilities and limitations of each GNN task when designing a model while also managing complexity and optimizing performance.

GNN has demonstrated a high potential for designing robust and dynamic models that solve many practical problems by executing one or more of the aforementioned tasks. In fact, GNN can execute different tasks using the same architecture. For instance, the DyGNN framework,~\cite{ma_streaming_2020}, was developed for both link prediction and node classification tasks applied to evolving graphs such as social networks and VSN. Furthermore, with GNN, various tasks can be applied to the same application domain to address its challenges. For instance, learning molecular fingerprints is a challenging task that primarily requires node classification, graph classification, and graph generation tasks~\cite{wu_comprehensive_2021}. Considering that, we can clearly conclude that complexity is always an issue that faces GNN models, especially if the framework is approaching more than one task, as in the above examples. Therefore, it is essential to keep in mind the different capabilities of each GNN task along with their features and limitations prior to designing the GNN model for the particular application domain and not to ignore minimizing the complexity as well as optimizing the final model.
\begin{figure}[tp]
    \centering
    \includegraphics[width=8.75cm]{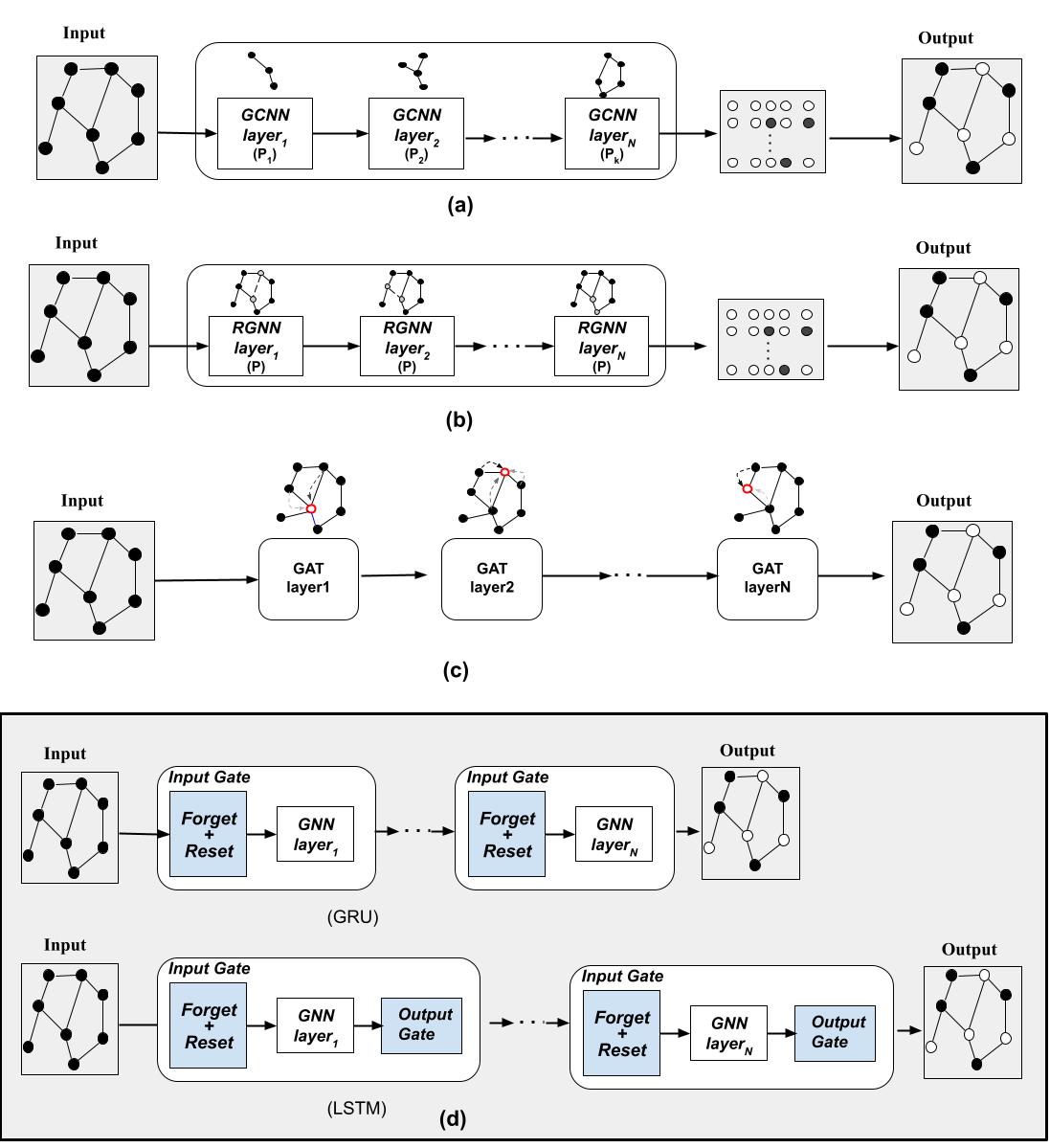}
    \caption{GNN general architectures where (a) represents the Graph Convolutional Neural Networks (GCNNs) which take the graph as input and update the hidden states of nodes using different parameters at each layer ($P_k$), (b) represents the Recurrent Graph Neural Networks (RGNNs) which learns message passing across nodes in the graph by recurrently updating hidden states using fixed parameters ($P$) in each layer, (c) presents the GATs with graph attention updating importance of neighboring nodes in each layer, and (d) presents the two Gated Graph Neural Networks approaches (Gated Recurrent Units (GRUs) and Long Short-Term Memory (LSTM)). }
    \label{fig:GNN types}
\end{figure}

\subsection{\textcolor{black}{GNN Main Architectures}}

Popular deep learning models like CNN, RNN, and autoencoders have influenced the development of GNNs~\cite{wu_comprehensive_2021}. The first generation recurrent GNN (RGNN), learned node representations by propagating neighbor information iteratively until convergence. Convolutional GNN (GCNN) emerged later, inspired by CNNs, to process graph data. Additionally, graph autoencoders were developed, leveraging the success of autoencoders for graph embedding tasks.


GNN architectures are highly dependent on data topology and application context. For example, modeling road traffic data requires considering graph type, data sources, environmental factors, and objectives. This section provides a concise overview of key GNN architectures, their technical aspects, advantages, challenges, and applications in VSN-related graphs. Fig.~\ref{fig:GNN types} illustrates their high-level structures.

\paragraph{Graph Convolutional Neural Network (GCNN)}
GCNN is a foundational GNN architecture known for effectively learning node representations and visualizing graph structures.  The core idea of GCNN is to produce a representation of the node ($V$)  by combining its own properties, say $X_v$, with neighbors' $X_u$, where $u \in N(v)$~\cite{wu_comprehensive_2021}. This is a fundamental step in many of the existing GNN frameworks, which are designed to solve problems using graph data structure. For example, the proposed DyGNN framework in~\cite{ma_streaming_2020} combines LSTM with GCNN to develop an architecture that can capture the dynamically changing graph structure, which can be seen in many real-world problems. Another illustration is the graph convolutional attention networks, proposed in~\cite{isufi_edgenets_2022} as one of the EdgeNet architectures with the goal of learning edge weights and convolution filter parameters. 

Unlike CNNs, GCNNs take both node and neighborhood features as input and apply stacked convolution layers with fixed-depth architecture. This design keeps parameter size independent of the graph size~\cite{ruiz_gated_2020}. GCNNs follow two main approaches:
\begin{itemize}
    \item \textit{The spectral-based approaches:} Uses graph signal processing to define filters and remove noise, assuming undirected graphs.
    \item \textit{The spatial-based approaches:} Updates node states using spatial relationships with neighbors, layer by layer, improving convergence and training speed. 
\end{itemize}
A comprehensive comparison between the two approaches and existing works is demonstrated in~\cite{wu_comprehensive_2021}.
Although GCNN can effectively model graph data structures, they are incapable of handling dynamically changing graphs especially the graphs that change over time or based on events. This limitation has been identified in a number of efforts, such as~\cite{ma_streaming_2020,manessi_dynamic_2020,gao_stochastic_2021}.

\paragraph{Recurrent Graph Neural Network (RGNN)}
\label{S:very_important}

RGNN is the most pioneering GNN. It aims to learn node representation by message passing/propagation technique. This technique assumes that each node in the graph is exchanging information with its neighbors continuously until it reaches a stable state. By using a predefined recurrent function with fixed parameters, RGNN generates a representation of these massages on recurrent bases to extract high-level node representations. The model receives a randomly initialized state of graph nodes $h_t^{(0)}$. Then, it updates the node hidden states by taking the current data point $X_t$ and the previous hidden states $h_{t-1}^{(n)}$ as inputs and generating the updated hidden state as an output from the current layer. The state of RGNN at any time $t$ is determined by parameterizing the linear mappings between linear maps $A$ and $B$ using a graph shift operator $S$.

While RGNN can efficiently capture temporal dependencies, it has limitations. Its iterative updates can be time-consuming compared to GCNNs, which use fixed-depth convolution layers~\cite{yu_spatio-temporal_2018}. Stability is another concern in RGNN. The graph structure affects how well RGNN performs. In other words, the output of RGNN would differ from what was anticipated if the graph changed or was estimated incorrectly~\cite{ruiz_gated_2020}. The proper design of the recurrent function is critical to ensure convergence and stable outputs~\cite{wu_comprehensive_2021}.. 

There are a wide number of real-world problem scenarios where RGNN can be an efficient solution. An example could be estimating earthquake origin region by learning from seismic wave data~\cite{ruiz_gated_2020}. In addition, RGNN is applied widely in the text analysis field, such as~\cite{wei_recurrent_2020}, which proposed a recurrent structure to aggregate the contextual information and update the hidden representation of the text. RGNN also has widespread applications in time series analysis.

\paragraph{Graph Attention Network (GAT)}
GATs~\cite{velivckovic2017graph}  extend traditional GNNs by using attention mechanisms to learn node representations. Unlike GCNs, which average the features of a node’s neighbors uniformly, GATs assign different importance to each neighbor using attention coefficients\cite{Wu_2021}. 
These coefficients are computed using a shared self-attention mechanism, which calculates an attention score for each pair of nodes. The scores are then normalized across each node’s neighborhood using a SoftMax function. The attention-based approach allows GATs to assign different weights to different neighbors, providing a more flexible and potentially more expressive model. It also offers a level of interpretability, as the attention coefficients can be seen as indicating the importance of each neighbor.  The parallel nature of attention computation across nodes makes GATs scalable and efficient for large graphs~\cite{zhang_graph_2022}. They have been successfully applied in tasks like traffic flow prediction~\cite{zhang_automatic_2022,zhang_fastgnn_2021}, optimizing traffic signals by ranking intersections~\cite{Zhong_probablistic_GNN_2021}, and solving combinatorial problems like the Traveling Salesman Problem (TSP)~\cite{nammouchi_generative_2020}.


\textcolor{black}{Recent work has also explored the synergy between GNNs and Transformers~\cite{vaswani2023attentionneed}. While GNNs are effective at local aggregation, Transformers excel at capturing global dependencies using self-attention. However, the quadratic complexity of Transformer attention can be a bottleneck in large-scale graphs. To address this, hybrid models combine GNNs with Transformers to balance local and global feature learning. For instance, Ruan et al.~\cite{9674238} proposed a GCN-Transformer hybrid to extract spatial-temporal features in traffic networks efficiently.}

\paragraph{Gated Graph Neural Network (GGNN)}

Traditional GNNs are effective when a single output is required (e.g., for node or graph classification), but they struggle with tasks that demand sequential outputs conditioned on time, space, or other contextual factors. Moreover, these models are susceptible to the vanishing gradient problem during training~\cite{ruiz_gated_2020}. Gated Graph Neural Networks (GGNNs) address these challenges by integrating gating mechanisms into the message-passing process.

Two primary variants of GGNNs exist: one based on GRUs and another on LSTMs units~\cite{zhou2021graph}. Both approaches are employed in the propagation step of the GNN to deal with the aforementioned limitations and enhance the propagation of the information over the whole graph structure~\cite{zhou2021graph}. GRU works by revealing a recurrence of GNN for a determined number of steps/recurrence $T$. It works by updating the current node's hidden states based on its prior hidden states~\cite{zhou2021graph}. This is accomplished by adding one gate, which concatenates a reset and input gates. The reset (forget) Gate $f$ decides if the information should be kept or forgotten. The input Gate $i$ specifies the amount of the currently processed state that the network has to keep. Those two gates are concatenated into one gate that is passed to the network function layer. On the other hand, the LSTM approach has an additional Output gate $o$ along with the input and forget gates. Its main purpose is to decide what the future hidden state should be. LSTM usually applies to a tree or graph structure GNNs, in which it is interesting to learn information about the children/siblings nodes as compatible with the parents nodes~\cite{ruiz_gated_2019}. The output gate allows LSTM to have another line of prediction by propagating information about the future nodes in the graph instead of updating based on the closely connected neighbors of the current node only.


There are two general approaches to building GGNN models: using a Gated Recurrent Unit (GRU) to the GNN function or applying a Long Short-Term Memory unit (LSTM). Both approaches are employed in the propagation step of the GNN to deal with the aforementioned limitations and enhance the propagation of the information over the whole graph structure~\cite{zhou2021graph}. GRU works by revealing a recurrence of GNN for a determined number of steps/recurrence $T$. It works by updating the current node's hidden states based on its prior hidden states~\cite{zhou2021graph}. This is accomplished by adding one gate, which concatenates a reset and input gates. The reset (forget) gate $f$ decides if the information should be kept or forgotten. 
The input gate $i$ specifies the amount of the currently processed state that the network has to keep. Those two gates are concatenated into one gate that is passed to the network function layer. On the other hand, the LSTM approach has an additional output gate $o$ along with the input and forget gates. Its main purpose is to decide what the future hidden state should be. LSTM usually applies to a tree or graph structure GNNs, in which it is interesting to learn information about the children/siblings nodes as compatible with the parents nodes~\cite{ruiz_gated_2019}. Output gate allows LSTM to have another line of prediction by propagating information about the future nodes in the graph instead of updating based on the closely connected neighbors of the current node only.  GGNN is applied in a wide range of applications including weather forecasting~\cite{Minbo_HiSTGNN_2022}, earthquake prediction~\cite{Ruiz_Spatial_Gating_2020}, traffic speed estimation~\cite{Xin_RealTime_traffic_2022}, short-term load of buses~\cite{Nantian_Gated_loadForecastingBuses_2023}, traffic prediction~\cite{Zhiyong_GGNN_traffic_2020}, and text analysis applications such as speech recognition models as well as text classification and summarization~\cite{Zeyu_Gated_abstractive_sum_2021, deng_text_2022}.


\subsection{Modeling VSNs for GNN}
Inspired by the wide range of GNN applications that have been discovered and reviewed in this section, we propose to explore their potential within the field of VSN. The complexity of traffic data makes it well-suited for GNN since they are able to capture the complex relationships and interactions among vehicles and other traffic participants. In other words, these data can be represented by generating graphs with various levels and architectural styles that are unique to the problems under the scope of VSN. Yet, implementing GNN algorithms for ITS and smart mobility applications using VSN graphs is still a growing field that has not reached maturity yet.
\begin{figure*}[htp]
    \centering
    \includegraphics[width=1\textwidth]{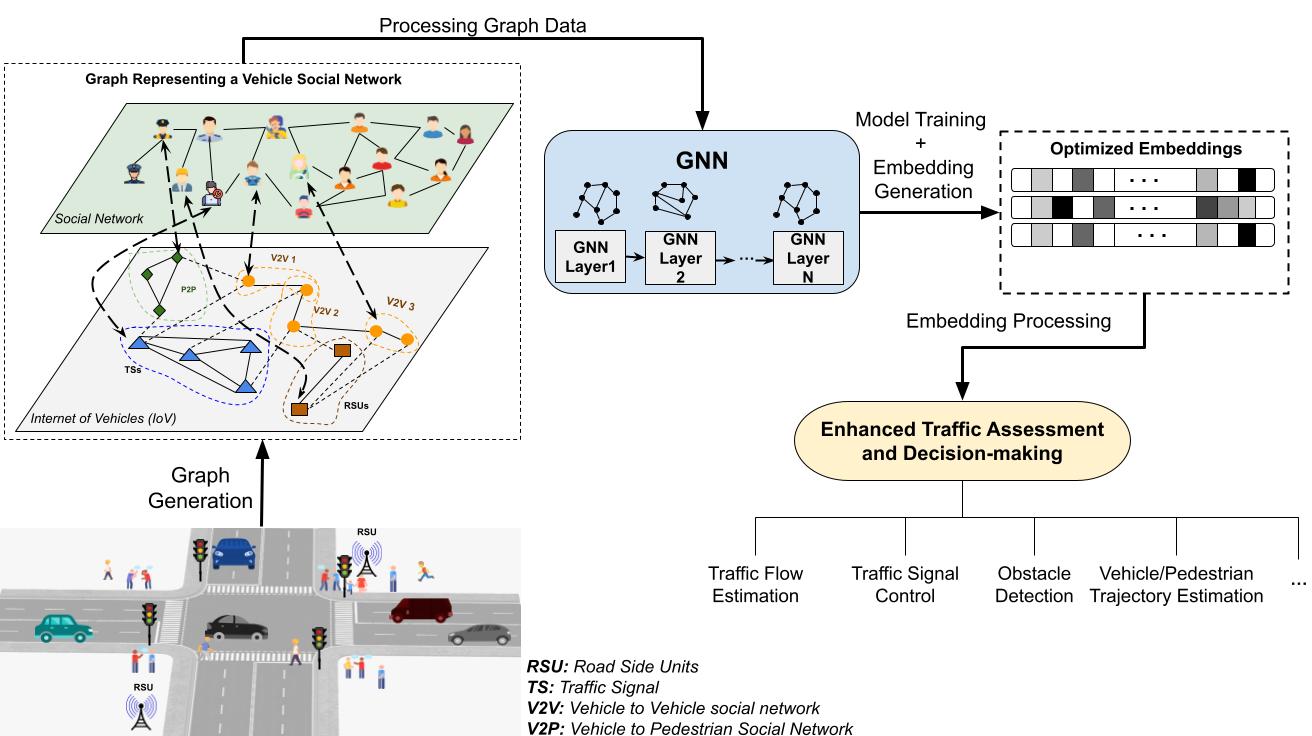}
    \centering
    \caption[width=10cm]{An illustration of applying GNN on an established VSN from a sample traffic scenario.}
    \label{fig:GNNonVSN_process}\vspace{-0.1cm}
\end{figure*}

Over the last decade, the main focus has been on designing conventional machine and deep learning models for different transportation problems. In~\cite{tingting_machine_2021}, the evolution of machine learning models in the ITS field is discussed. The survey provides a comprehensive overview of the machine/deep learning techniques and methods that have been used for different ITS applications, such as traffic prediction, traffic control, and transportation mode recognition. Since VSN-related applications involve modeling and understanding interactions among vehicles or between vehicles and other traffic participants, VSN can be considered a fundamental subset of ITS.

Investigating vehicular and social networks in ITS has recently attracted much attention, especially when the transportation and mobility problems are modeled as graph-structured data. When designing GNN models for VSN applications, it is crucial to consider the following factors:
\begin{itemize}
    \item \textit{Structure of the graph:} The GNN model must be able to effectively understand the interactions between the different actors, their features, and their interactions in the system. The graph structure should be able to capture the different relationships between the nodes reflected through the edges.
    \item \textit{Spatial/temporal dependencies:} In many applications, such as trajectory prediction, it is important to model the temporal and spatial dependencies of the users, including their movements and speeds, as well as their entrance and exit from the network.
    \item \textit{Training and Evaluation:} The GNN model needs to be trained and evaluated on suitable datasets that capture the relevant factors (i.e., the distribution of the nodes and edges as well as their features) of the system so that patterns in the data can be identified.
    \item \textit{Complexity:} The GNN model complexity needs to be taken into account in terms of computation and memory usage, especially for real-time applications.
\end{itemize}
Fig.~\ref{fig:GNNonVSN_process} presents a suggested general approach to modeling dynamic VSN as graphs for the GNN model. Individual objects such as vehicles or road users can be represented as nodes, and interactions between those objects (e.g., social vehicular relations) as edges. This can guide the model in capturing a range of useful data such as speed, traffic patterns, etc. The generated graph or set of graphs can then be fed into the selected GNN model, which aggregates data from every node's neighbors over several layers to process the graph. The model learns the optimized embeddings that better encode the spatial and temporal aspects of the network. The resulting high-level representations of the VSN entities and interactions provided by these optimized embeddings facilitate the assessment and analysis of the traffic patterns. This leads to better traffic understanding and real-time decision-making procedures in various applications, which we will discuss in detail in the next section.

\begin{table*}[]

\caption{Examples of the GNN models proposed in the surveyed studies and applied to the VSN-related tasks}
\label{tab:table2_v2}
\centering
\begin{tabular}{|c|c|c|c|}
\hline
\textbf{Tasks}                                              & \textbf{Ref.}                                             & \textbf{Proposed Model}               & \textbf{GNN Variation(s)}          \\ \hline
\multirow{5}{*}{Flow Prediction}                            &~\cite{chen_aargnn_2022}          & AARGNN                                & GRNN + LSTM                        \\ \cline{2-4} 
                    &\textcolor{black}{~\cite{iTPGT_former_2024}}
                    &\textcolor{black}{iTPGT-former}
                    &\textcolor{black}{GAT + Graph Transformer}
                    \\ \cline{2-4}
                    & ~\cite{zhou_variational_2021}     & VGRAN                                 & Hybrid  (GCNN, GRNN, GRU)          \\ \cline{2-4} 
                    
                    & ~\cite{bui_spatial-temporal_2022} & STGNN                                 & Hybrid  (GCNN, GRNN)               \\ \cline{2-4} 
                    
                    & ~\cite{zhang_automatic_2022}      & GDN                                 & Hybrid (GCNN, GATs)                \\ \cline{2-4} 
                    
                    & ~\cite{li_spatial-temporal_2021}  & STFGNN                                & Hybrid  (GCNN, GGNN)               \\ \hline
\multirow{4}{*}{Traffic Forecasting}                        & ~\cite{bai_adaptive_2022}         & AGCRN                                 & Hybrid (GCNN, GRU)                 \\ \cline{2-4} 
                                                            & ~\cite{zhang_fastgnn_2021}        & FASTGNN                               & Hybrid (GATs, GRU)                 \\ \cline{2-4} 
                                                            & ~\cite{tian_st-mgat_2020}         & ST-MGAT                               & Hybrid (GATs)                      \\ \cline{2-4} 
                                                            & ~\cite{zhang_graph_2022}          & Gra-TF                                & Hybrid (GCNN, GATs)                \\ \hline
\multirow{3}{*}{Traffic Signal Control}                     &~\cite{Zhong_ProbGNN_2021}        & TSC-GNN                               & Hybrid (GCNN, GATs)                \\ \cline{2-4} 
                                                            & ~\cite{Li_DeepImitation_2020}     & Deep imitation learning model         & GCNN + LSTM                        \\ \cline{2-4} 
                                                            & ~\cite{Wang_STMARL_2022}          & STMARL                                & GRNN + Reinforcement Learning      \\ \hline
\multirow{2}{*}{Driving Assistance and Autonomous Vehicles} 
                                                            & ~\cite{ha_road_2021}              & Road-GNN                              & GCNN                               \\ \cline{2-4} 
                                                            & ~\cite{jin_graph_2022}            & APID-Net                              & GCNN     
                            \\ \hline
Pedestrian Recognition, Obstacle Detection                  & ~\cite{zhou_ast-gnn_2021}         & AST-GNN                               & GCNN                               \\ \hline
\multirow{3}{*}{Trajectories Prediction}                    & ~\cite{mo_graph_2021}             & GNN-RNN based Encoder-Decoder network & Hybrid (GATs, GGNN)                \\ \cline{2-4} 
                                                            & ~\cite{chen_robust_2021}          & Toast Framework                       & GRNN                               \\ \cline{2-4} 
                                                            & ~\cite{li_hierarchical_2022}      & hierarchical GNN framework            & GCNN                               \\ \hline
Vehicles Routing Problems (VRP)                             & ~\cite{nammouchi_generative_2020} & GLN-TSP                               & GCNN + GATs                        \\ \hline
\multirow{2}{*}{Traffic Data Analysis}                      & ~\cite{liu_tap_2022}              & TAP                                   & GCNN + AutoEncoders                \\ \cline{2-4} 
                                                            & ~\cite{guo_mixed_2022}            & Mixed-GNN                             & Hybrid  (GCNN, GRNN)               \\ \hline
\multirow{3}{*}{Connectivity and Computing}                 &~\cite{10509567}                    & MDP-GAT-DRL                           & GCNN + Deep Reinforcement Learning \\ \cline{2-4} 
                            & ~\cite{GNN_joint_comm}                    & GCNN-JCAS                             & Heterogeneous GCNN                 \\ \cline{2-4} 
                                                            & ~\cite{10637671}                    & Transedge                             & GRNN + Deep Reinforcement Learning \\ \hline
\end{tabular}

\end{table*}


\section{Applications of GNN on VSN}
\label{sec4}
This section comprehensively reviews the main VSN-related applications that implement different GNN models. We have compiled recent research efforts, mainly from January 2020 to September 2024, utilizing GNN architectures and categorized them according to key ITS applications, namely traffic flow prediction, trajectory prediction, traffic forecasting, traffic signal control, driving assistance and autonomous vehicles, vehicle routing problems, traffic data analysis, and connectivity and computing in ITS. In selecting studies for review, we focused on ITS applications where input data is modeled as graphs. Importantly, we only included studies where the investigated graphs originated from a VSN or a subgraph of a VSN. In other words, we reviewed studies where the graphs were derived from an IoV, VANET, social network, or a combination of these. Studies that do not involve a network or connected objects were excluded from this survey. For instance, a study where the graph is constructed solely from the perception of a single-ego vehicle is not considered. 

\begin{figure}[tp]
    \centering
    \includegraphics[width=0.5\textwidth]{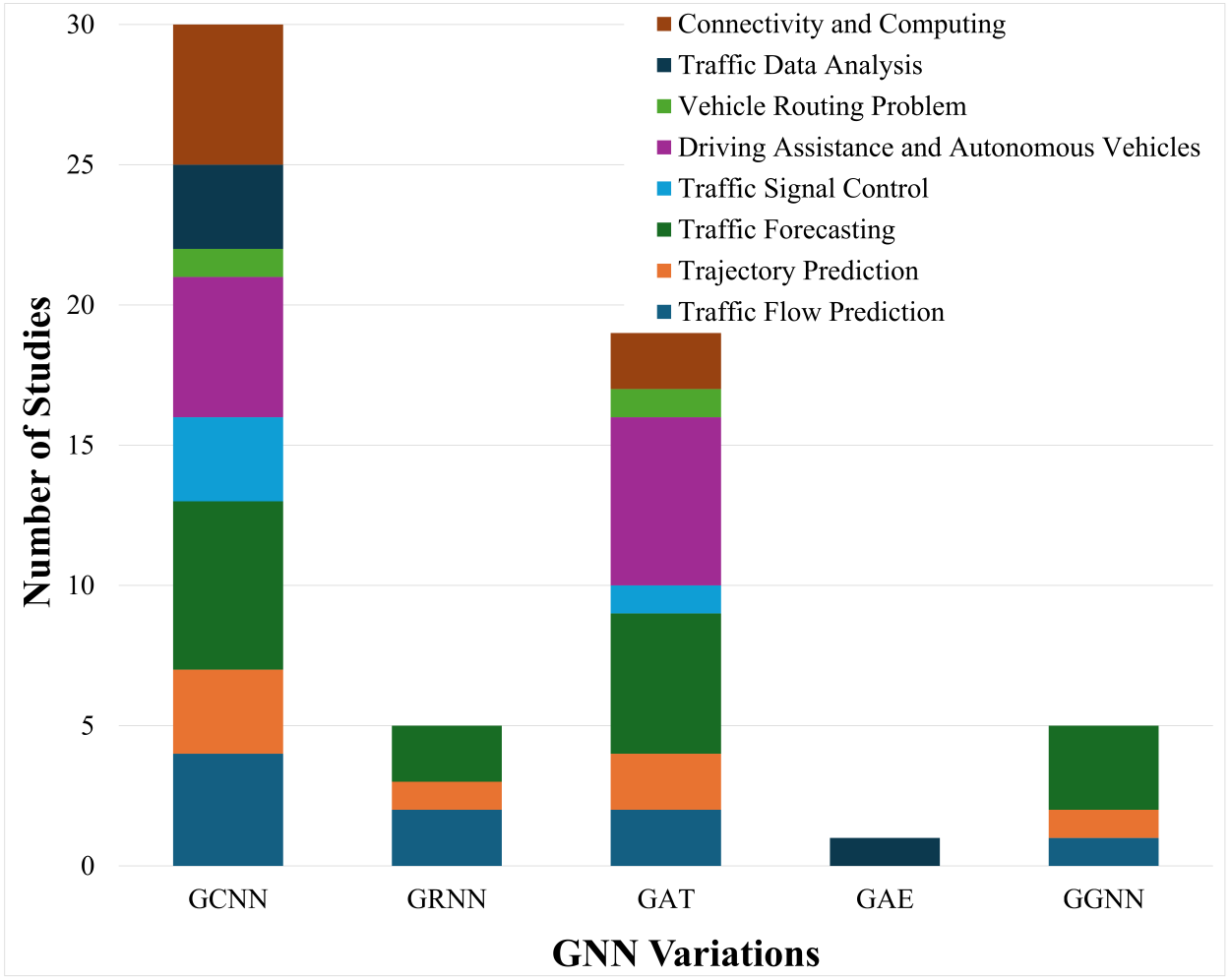}
    \centering
    \caption[width=10cm]{\color{black}A bar graph illustrating the distribution of different implementations of GNN architectures across various categories of ITS applications on VSN-related graphs. The x-axis represents GNN architectures (GCNN, GRNN, GAT, and GGNN), while the y-axis shows the number of studies that employed each type. Hybrid models employing multiple architectures are counted in each relevant category.}
    \label{fig:gnn_vsn_stat}\vspace{-0.1cm}
\end{figure}

Table~\ref{tab:table2_v2} provides examples of the collected studies within each ITS application, highlighting the specific GNN models utilized. It reflects the diversity of the proposed GNN models in the literature and categorizes them according to the main utilized GNN architecture. We noticed that many of the proposed models involved a combination of GNNs leveraging the advantages of both architectures. For instance, we can mention the Spatial-Temporal Fusion Graph Neural Networks (STFGNN) in~\cite{li_spatial-temporal_2021} that exploit the convolutional and recurrent aspects of GNNs to enable effective flow prediction. Other studies proposed to combine GNN architectures with reinforcement learning techniques when decision-making is needed, such as in the context of smart traffic control. The distribution of the surveyed studies that utilize each GNN architecture is presented in Fig.~\ref{fig:gnn_vsn_stat}, which are categorized based on the application area. The figure presents a stacked bar graph comparing the use of different GNN architectures, GCNN, GRNN, GAT, and GGNN, across various Intelligent Transportation System (ITS) applications. We can observe that GCNN shows the highest number of studies, followed by GAT, with applications across multiple ITS tasks. GRNN and GGNN are used less frequently, primarily in traffic Forecasting and traffic flow prediction, as these applications require considering the temporal aspects of the data. This comparison highlights the widespread adoption of GCNN and GAT for diverse ITS applications applied on VSN or their sub-component graphs. It is important to note that applying GNN models to VSN-related tasks is still in its early stages, requiring further research and experimentation, as relatively few studies have focused on a holistic VSN to date.

\begin{figure*}[tp]
    \centering
    \includegraphics[width=1\textwidth]{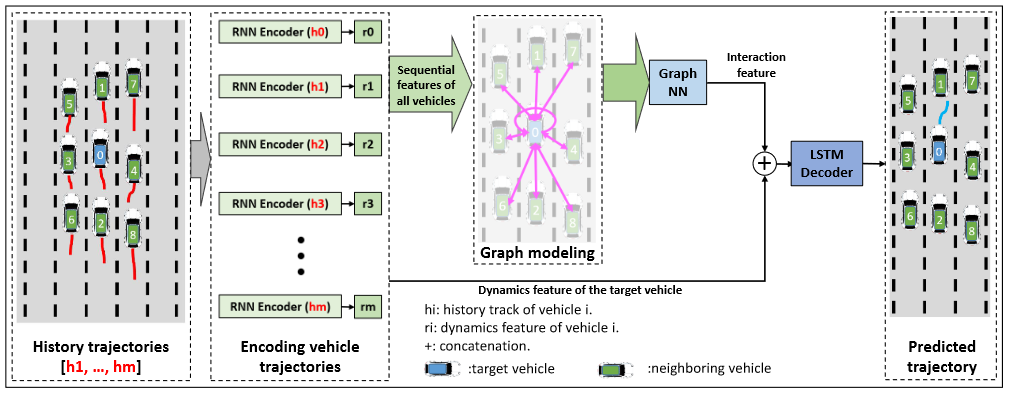}
    \centering
    \caption[width=10cm]{Illustration of the proposed model adopted from~\cite{mo_graph_2021}. RNNs with shared weights are used to encode the dynamics features of vehicles individually. A GNN-based interaction encoder is applied to these dynamics features, which are contained in corresponding nodes in a directed interaction graph, to summarize the inter-vehicular interaction feature. Finally, an LSTM decoder predicts the trajectory by jointly considering the target vehicle’s dynamics and interaction features.}
    \label{fig:Trajectory_GNN_example}\vspace{-0.1cm}
\end{figure*}
\subsection{Traffic Flow Prediction (TFP)}

Predicting a flow pattern is about estimating an accurate expected traffic flow behavior based on the current data about roads, vehicles, relations between vehicles, and vehicles to other road users. Vehicles can form a social network to share important information about their surroundings in real-time, which can help in learning to perform various tasks. Examples include vehicle trajectory prediction, anomaly detection, traveling time estimation, and prevention of potential accidents. The aim is to optimize traffic management tasks, such as mitigating congestion, reducing traveling time, and monitoring the transportation network. It can also support the development of full and partial autonomous vehicles by providing the appropriate estimations that can greatly enhance the system's intelligence~\cite{10364647}. With the increasing amount of connected vehicles on the road and the growth of real-time data, designing robust flow prediction models is becoming more important than ever. However, this task is not trivial due to the dynamically changing spatial and temporal parameters and the changing factors of other road commuters. With GNN, this type of task can be modeled by graphs with different road entities as nodes and their connection or type of relations selected as the edges. Relations can also be classified based on the scope of the problem domain. Regardless of the design and architecture of the GNN model, the target should always be to reduce the complexity of the generated model as much as possible and find the best generalized architecture that can capture the desired patterns. 

Few studies introduced GNN models to optimize the flow prediction task. The authors of~\cite{chen_aargnn_2022} presented a new approach for traffic flow prediction using GNNs. They proposed an attentive attributed recurrent GNN that considers multiple dynamic factors, including spatial-temporal correlations, traffic flow dynamics, and historical traffic data. The model uses an attention mechanism to weigh the importance of different factors in the prediction and an attributed graph to model the interactions between traffic flow and factors such as weather conditions and holidays. In~\cite{zhou_variational_2021}, the authors proposed a variational GNN, a road traffic prediction model that employs GNNs to learn the underlying representations of the traffic network and uses variational inference to model the uncertainty in traffic predictions. The model uses a convolutional GNN to extract the spatial-temporal features from the traffic data and a variational recurrent neural network to capture the temporal dependencies and predict the traffic's future flow. This research field is still new and requires investigation of more solutions using GNNs.

\begin{figure}[tp]
    \centering
    \includegraphics[width=8cm, height=5cm]{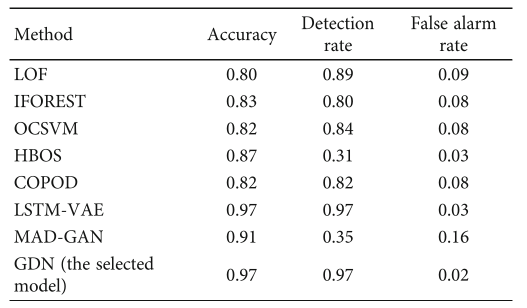}
    \vspace{-0.3cm}
    \caption[width=8.75cm]{\color{black}Table comparing the performance of the proposed GNN-based model (GDN) in~\cite{zhang_automatic_2022}, with other non-GNN models for traffic flow prediction application. The higher numbers represent better performance.}
    \label{fig:Traffic_Prediction_results}\vspace{-0.4cm}
\end{figure}

Capturing the spatial and temporal factors in road traffic can greatly impact the traffic flow and pattern prediction tasks. It has shown to be very effective in providing detailed information about real-time traffic and improving the accuracy of GNN models. This approach led to an emerging era in the flow prediction field~\cite{bui_spatial-temporal_2022}. Several proposed models captured those factors with Spatio-Temporal-GNN (ST-GNN) models. For example, the study in~\cite{zhang_automatic_2022} presents an ST-GNN model for detecting traffic anomalies on a road network using data from infrastructure-based traffic sensors to learn a representation of the traffic pattern. The learned representation helps detect anomalies in the traffic data, such as sudden changes in traffic volume or unusual congestion. {\color{black}Fig.~\ref{fig:Traffic_Prediction_results} demonstrates the performance of the GNN-based model (GDN) proposed in~\cite{zhang_automatic_2022} compared with the baseline. The table shows that the GNN-based model outperforms the compared models, presenting the effectiveness of GNN on traffic flow prediction tasks.} In addition, the authors of~\cite{li_spatial-temporal_2021} were able to model the interactions between different traffic nodes and also incorporate historical traffic data to make predictions about future traffic flow. The authors of~\cite{bui_spatial-temporal_2022} provided a taxonomy for the ST-GNN models and a general model architecture with a comprehensive review of the existing solutions proposed using different GNN architectures. It is shown that measuring the spatial and temporal characteristics is not a trivial task to achieve, as most of the existing solutions either suffer from high complexity or generalization.
In recent work~\cite{STGNN_IoV}, LSTM and GRU are integrated to estimate traffic flow, considering spatial relationships in the road network and temporal traffic data patterns. The graph is constructed by representing roads as nodes and their features, such as volume, average speed, etc. The connection between roads creates edges in the graph that define how traffic from one road influences the traffic to the other road. An example is to capture the traffic flow between adjacent road sections, which then helps model the traffic flow across the road network.

\subsection{Trajectory Prediction (TP)} 
Trajectory prediction can fall under flow prediction tasks, as it requires analysis of objects' movement in real-time within a relatively short time period. It aims to predict the future movement of vehicles or other road users as well as their interaction with other objects until they reach their destination. This task is not a trivial task because it requires the model to learn the social interactive behavior of objects within the environment and the neighboring objects, which could be continuously moving or changing. Therefore, the graph representation of the different VSNs established in the traffic shows great potential in solving this problem. The authors of~\cite{mo_graph_2021} presented a model for predicting the trajectories of vehicles on a highway using a combination of GCNN and RNNs. As illustrated in Fig.~\ref{fig:Trajectory_GNN_example}, the model uses GCNNs to model the interactions between vehicles and the highway infrastructure and RNNs to model the temporal dependencies of the vehicle trajectories.

The dynamic and complicated traffic situations that exist in such environments increase the challenges to ensuring a high-level of safety in trajectory prediction. That makes road network representation learning one of the highly demanding tasks in VSN, as it requires effectively capturing both the traffic patterns as well as the traveling semantics.In this regard, the authors of~\cite{Vehic_Inter} proposed a dynamic GNN model that captures the complex spatial and temporal interaction between vehicles, distributions, and varying speeds, as well as their relation with the road network topology using proximity-based timestamp graphs construction mechanism on a real-world vehicle's data. The model proposed in~\cite{AMGB}, on the other hand, integrates GNN with BiLSTM and incorporates an attention mechanism in order to effectively predict vehicles' directions as well as the motion distances between vehicles. The graph is constructed by modeling every single vehicle as a node that holds multimodal features such as speed, position, and type. The edges represent the interactions between vehicles, which is based on spatial proximity. That means edges constructed between vehicles that are near each other to influence the interaction among vehicles by sharing their movement features. Another proposed approach tackles the issue of missing trajectories of offline vehicles in the VSN environment~\cite{deep_traject}. The model captures vehicle trajectories by learning various driving behaviors and reconstructs missing trajectories by applying spectral graph analysis.

\begin{figure}[tp]
    \centering
    \includegraphics[width=7.5cm]{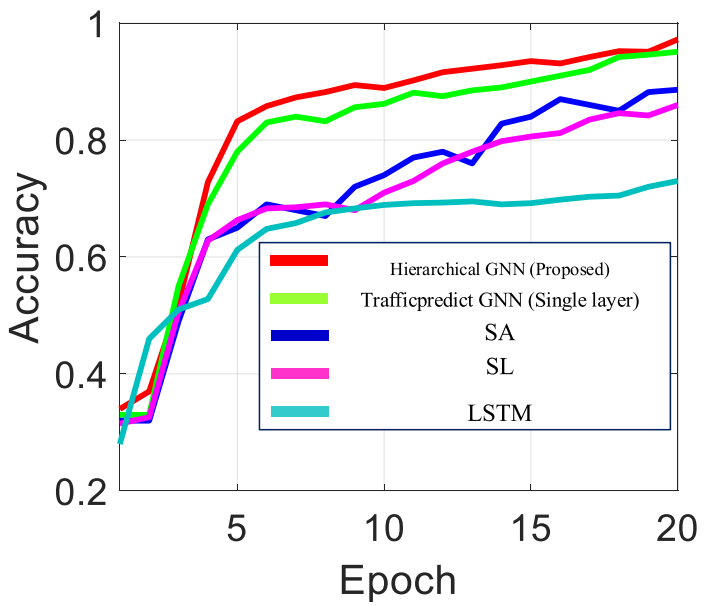}
    \vspace{-0.3cm}
    \caption[width=8.75cm]{\color{black} A comparison of training accuracy for trajectory prediction presented in~\cite{li_hierarchical_2022}. The graph shows GNN models (Hierarchical GNN and Trafficpredict GNN) with higher accuracy compared to other non-GNN models.}
    \label{fig:trajec_pred_result}\vspace{-0.4cm}
\end{figure}

For instance, the proposed solution in~\cite{chen_robust_2021} presented a method for learning robust representations of road networks, which can be utilized for various transportation-related tasks, such as traffic prediction and route planning. The method uses a combination of GNNs and unsupervised learning techniques to learn representations of road networks that are robust to variations in traffic patterns. On the other hand, the authors of~\cite{li_hierarchical_2022} presented a framework for predicting the possible interactive behavior of heterogeneous traffic participants (e.g., vehicles, bicycles, pedestrians) based on a GNN approach. The framework is hierarchical in nature, consisting of two levels: a global level and a local level. At the global level, the GNN model captures the interactions between different traffic participants and the road infrastructure in the form of a graph. At the local level, the GNN model captures the interactions between traffic participants in a specific area.
{\color{black}Fig.~\ref{fig:trajec_pred_result} presents the promising performance of GNN in traffic prediction compared to baseline methods. It shows that GNN-based solutions outperform non-GNN-based solutions by achieving higher accuracy in trajectory prediction.} This field is still developing and very demanding because it can serve as a foundation for other VSN tasks, such as abnormal driving detection and accident prediction.

\subsection{Traffic Forecasting (TF)}
\begin{figure}[tp]
    \centering
    \includegraphics[width=8cm, height=6cm]{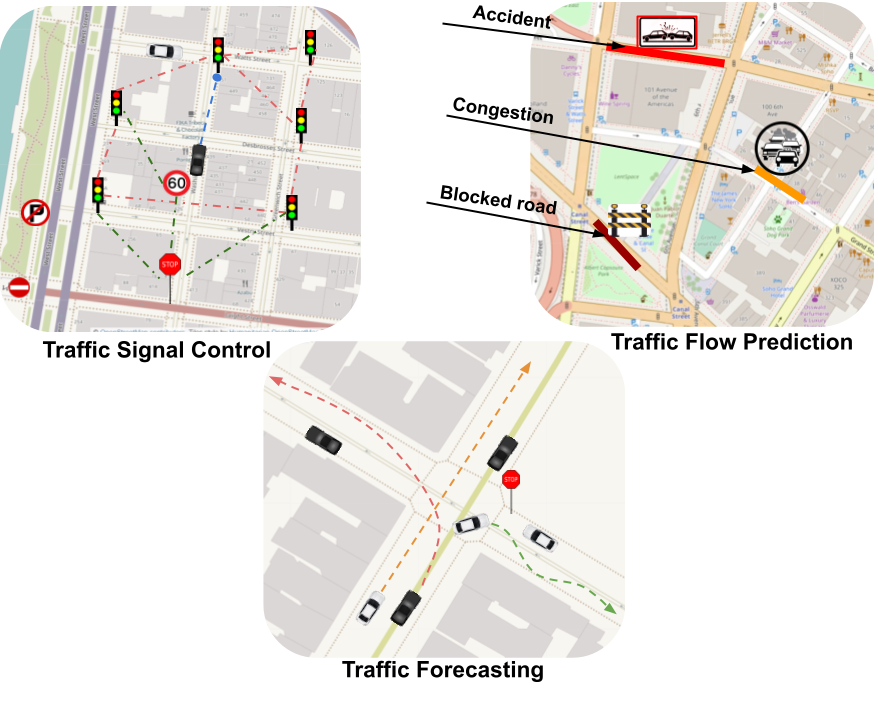}
    \vspace{-0.5cm}
    \caption[width=8.75cm]{Visualization of the traffic signal control, the traffic forecasting, and the traffic flow prediction.}
    \label{fig:TPR_Tasks}\vspace{-0.4cm}
\end{figure}

Although VSN can be established while the vehicle is within a defined region close to other vehicles, infrastructure, or road users, the amount of useful information shared over time and in different regions can help in learning to make a robust analysis of traffic patterns in the city and take impactful decisions. In traffic forecasting, the main goal is to predict future insights about traffic on road networks~\cite{bai_adaptive_2020}. It depends mainly on the current historical data about traffic, along with external factors that can affect the traffic status. For example, the weather, seasons, school calendars, etc.~\cite{jiang_graph_2022}. Fig.~\ref{fig:TPR_Tasks} illustrates an example of traffic forecasting problems to predict the future path of the vehicles. To generate graphs of road networks, road intersections can be represented as nodes and roads as edges. The data captured in traffic forecasting could be information about past traffic patterns, weather conditions, and road closures. Traffic forecasting can be divided into several categories, depending on the specific focus of the forecast. For example, some traffic forecasts may focus on predicting traffic flow on a particular road or intersection, while others may focus on predicting the demand for transportation services in a specific area. Additionally, traffic forecasts can be short-term or long-term, depending on the time frame over which they are predicting traffic patterns. Overall, the goal of traffic forecasting is to provide valuable information that can be used to improve the efficiency and effectiveness of transportation systems.

FASTGNN is one of the novel GNN-based approaches applied in traffic forecasting~\cite{zhang_fastgnn_2021}. It uses federated learning to improve the traffic speed forecasting task and graph sampling to partition the graph data into several sub-graphs, which are then distributed to different edge devices for training. Then, a novel method called Topological Information Protection (TIP) is applied to ensure that the topological information of the graph is not leaked during the training process. In addition, the authors of~\cite{bai_adaptive_2020} used a combination of Graph Convolutional Networks (GCNs) and Recurrent Neural Networks (RNNs) that adapt the graph structure during the training process to improve the performance of the model. GATs could also empower the forecasting models significantly. For instance, the work in~\cite{tian_st-mgat_2020} presented a Spatial-Temporal Multi-Head GAT (ST-MGAT), which uses multiple attention heads to capture both spatial and temporal dependencies in the graph data. Another example is the model proposed by~\cite{zhang_graph_2022}, which utilizes historical traffic data and real-time sensor data to predict traffic conditions on a given road network. Recent work proposes a graph multi-attention-based model that combines node-level attention with spatial and temporal attention to enhance the prediction of complex traffic networks~\cite {PGSLM}. They construct the graph by modeling roads and traffic sensors as nodes, and the relationships between these objects are the edges weighted by the Euclidean distance between them.

\begin{figure}[tp]
    \centering
    \includegraphics[width=8cm, height=6cm]{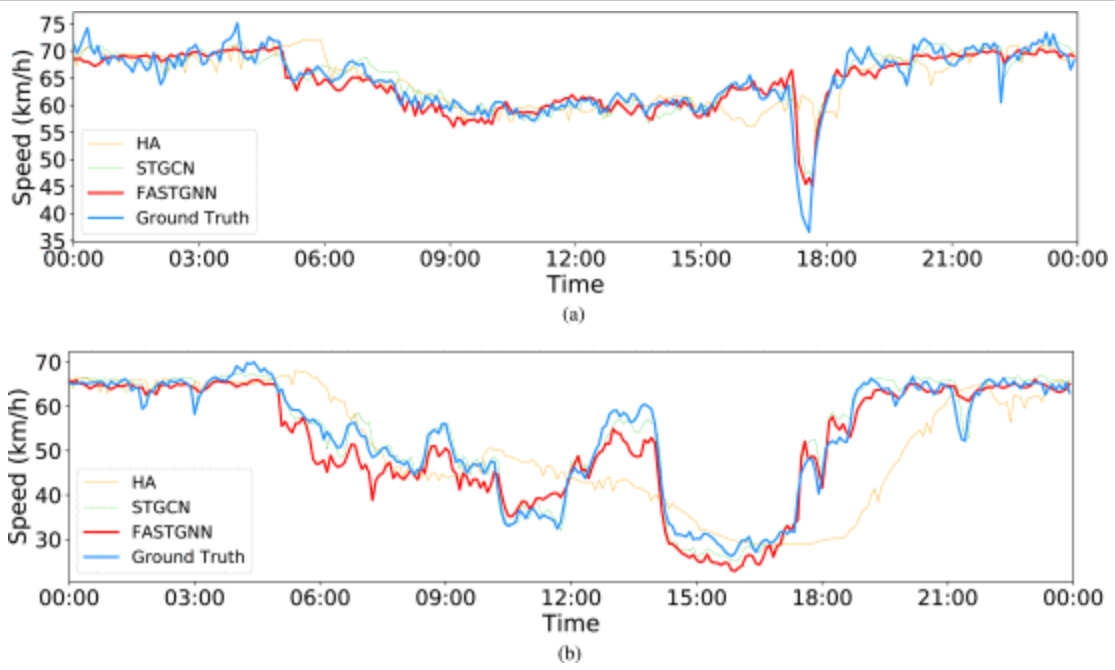}
    \vspace{-0.3cm}
    \caption[width=8.75cm]{\color{black} Traffic speed forecasting performance, reported in~\cite{zhang_fastgnn_2021}, showing the improved performance in FASTGNN and STGCN with ground truth data, compared with a non-GNN model, Historical Average approach (HA).}
    \label{fig:Traffic_Forcasting_results}\vspace{-0.4cm}
\end{figure}

One of the challenges when using GCNN to capture the spatial aspects of the graph data is that the graph structure is fixed during the training process, which means that nodes, edges, and parameters are not changed~\cite{wenjuan_research_2022}. This does not reflect the traffic topology in practice, which is continuously changing. That encourages researchers to support using RGNN as part of the model architecture, as it can take into account historical traffic patterns as well as the complex interactions between different traffic nodes, such as intersections and roads. In addition to RGNN, the attention mechanism can significantly improve the model by selectively weighing the importance of different graph components, which can be particularly useful for traffic forecasting, where traffic patterns can vary greatly between different parts of the graph. {\color{black} Fig.~\ref{fig:Traffic_Forcasting_results} shows the improvement in speed forecasting performance with STGCN and FASTGNN models compared to the Historical Average (HA) approach, with ground truth data~\cite{zhang_fastgnn_2021}.}

\subsection{Traffic Signal Control (TSC)}

Since vehicles in VSN can share real-time data such as location, speed, directions, and their intended route, traffic signals may use these data to optimize signal control by predicting traffic patterns and congestion and adjusting signal timing dynamically based on the specific traffic conditions rather than using pre-scheduled timing.
This task aims to optimize traffic flow by reducing travel time and decreasing fuel consumption through optimizing the control of traffic signal lights (an example in Fig.~\ref{fig:TPR_Tasks} shows how a graph may connect signals to enhance managing their operation). GNN can be used to understand the relations and interactions between traffic signals, road networks, and traffic flow. One approach to constructing a graph from a road network is to represent each traffic signal as a node, and the roads connecting the signals are represented as edges. Graph Convolutional layers can then be used to learn the relationships between the nodes and edges in the graph structure to perform the desired task. In addition, it can process real-time traffic data and update the hidden state of each node in the graph, which represents the state of each traffic signal.

The authors of~\cite{Zhong_ProbGNN_2021} and~\cite{Li_DeepImitation_2020} presented hybrid GNN models that combine GCNN with advanced learning approaches. In~\cite{Zhong_ProbGNN_2021}, combined probabilistic learning algorithms to better model the complex relationships and patterns between traffic flow, road networks, and traffic signals. The authors of~\cite{Li_DeepImitation_2020} proposed an imitation Learning-based model by optimizing the time of traffic signals, such as how long each signal should stay green or red. The STMARL model proposed in~\cite{Wang_STMARL_2022} combines reinforcement learning with a multi-agent system based on GNN to optimize the control decisions by capturing the interactions between signals and coordination between signal timings. It is essential to assess the performance of GNN-based traffic signal control methods with real-time data to ensure that they can produce informed decisions about traffic signal timing and adapt to changes in traffic patterns due to environmental and human factors.

\subsection{Driving Assistance and Autonomous Vehicles (DA/AV)}
The problem of driving assistance has been widely investigated in the literature with the development of advanced driver-assistance systems (ADASs); however, much more research effort has been invested to enable enhanced driving assistance with the emergence of autonomous vehicles. According to the current research, integrating driving assistance and autonomous vehicles is fundamental to guaranteeing high-standard levels of safety, efficiency, and sustainability. Therefore, there is a high demand to build robust deep-learning models that can maximize those potential benefits. \textcolor{black}{In this context, VSN provides a crucial communication framework by establishing dynamic networks among vehicles, roadside units, traffic infrastructure, and other entities, enabling real-time information exchange and cooperative decision-making. GNNs, on the other hand, leverage their ability to learn from structured graph-based representations of complex vehicular interactions and serve as a key enabler in several tasks, such as object detection and tracking, scene understanding, predictive modeling, trajectory planning, sensor fusion, and many more. Since this is a newly developing area in automotive research and industry, we present some promising solutions to support VSN in autonomous driving applications.} {\color{black}For example, a novel Road-GNN model has been proposed and simulated in an environment that is unseen by the model in the training phase~\cite{ha_road_2021}.} This model shows that graph representation learning can become a game-changer in this field by enhancing the self-driving car's performance and taking it to the next level. Another application in this area is detecting abnormal perception information to ensure safe driving. Another solution has been proposed by~\cite{jin_graph_2022} to detect abnormal information that could be received by self-driving vehicles through learning the behavior of the surrounding objects, which are also affected by their neighbors. This type of graph-based structured data can be effectively modeled using GNN, as presented in this paper.


In a recent work~\cite{wang2024rs2g}, authors developed a data-driven approach to model traffic scenarios as graphs utilizing a Transformer-based edge encoding technique to capture the complex relationships among road users. {\color{black}Similarly, the authors of~\cite{multi_modal_trans} proposed a multi-head attention Transformer encoder to extract relations among interacting vehicles to predict their behavior, which helps enhance the safety and efficiency of autonomous driving.}
Another work proposes a novel Stochastic GNN model that enhances the multi-agent reinforcement learning to capture the time-varying topological relations between connected vehicles~\cite{xiao2023stochastic}. Since VSN enables the vehicle to share real-time data about its speed, direction, car health, and the surrounding environment, this helps optimize tasks like cooperative driving and, therefore, enhances driving safety. For example, when a vehicle merges into a highway or wants to cross a road intersection, it can cooperate better with the other vehicles to make this merging smooth and avoid sudden breaks or changing lanes.

Another main task in this field is anomaly detection and obstacle avoidance. Establishing a combination of social networks among vehicles and other road users can enhance the vehicle's knowledge about any human crossing the road or a sudden obstacle in real-time, which can enhance traffic safety significantly. An example effort in this area is~\cite{zhou_ast-gnn_2021}, which presents a novel interaction-aware trajectory prediction model (AST-GNN) for predicting pedestrian trajectories in crowded environments. The authors used a spatiotemporal graph to visualize the time relationships between pedestrian interactions and an attention mechanism to evaluate the significance of various pedestrian interactions. This shows how GNNs are well-suited for solving VSN-related problems in such application domains. For instance, it can effectively process data captured from LiDAR sensors, which can be utilized to detect obstacles around the vehicle. It is worth mentioning that there is no GNN model that has been designed specifically for obstacle detection in VSN-related tasks. However, there is a range of GNN models that address the object detection task, such as~\cite{Thakur_GraphAttention_2022,Shi_PointGNN_2020,He_Svga_net_2022,qian_badet_2022}.


\begin{figure}[tp]
    \centering
    \vspace{-0.2cm}
    \includegraphics[width=8cm]{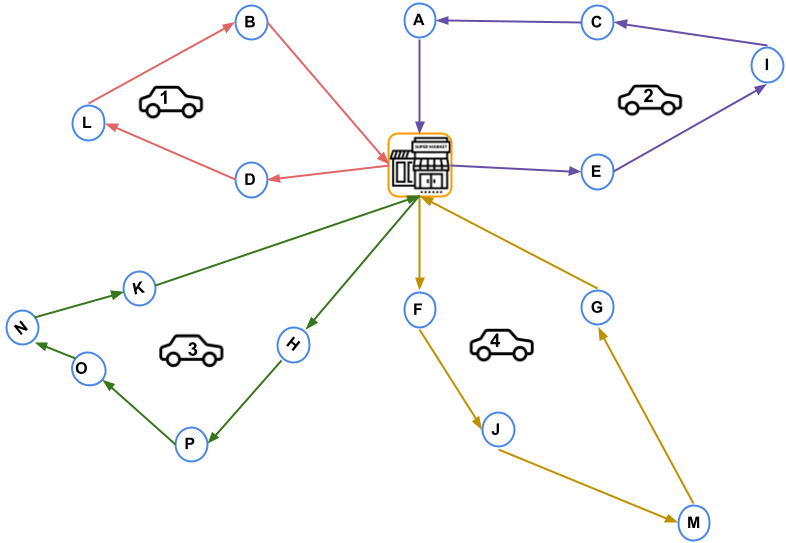}
    \caption[width=10cm]{A sample demonstration of the Vehicles Routing Problem (VRP).}
    \label{fig:Vehicles_routing_Problem}\vspace{-0.5cm}
\end{figure}

\subsection{Vehicle Routing Problem (VRP)}

Vehicle Routing Problems (VRP) are combinatorial optimization problems in which a fleet of vehicles must be assigned routes to deliver goods to a set of customers. The goal is to minimize the total cost of the routes, which includes the cost of travel, the cost of loading and unloading goods, and the cost of waiting time. Many VRP-related problems can be derived from the original problem depending on the objectives and constraints considered. Examples are the capacitated VRP, where limited capacities of the goods are imposed for each vehicle, or the VRP with time windows, where each delivery location is characterized by a given time window within which the deliveries must be made. VRP, as depicted in Fig.~\ref{fig:Vehicles_routing_Problem}, is considered one of the most challenging problems to solve, and there is no known efficient algorithm that can find the optimal solution for all instances of the problem. However, there are a number of heuristics that can be used to find good solutions, such as the nearest neighbor algorithm according to the Clarke-Wright algorithm~\cite{Herdianto_Guided_2021}. We observed that most of the current solutions to this problem are based on static data related to traffic roads and congestion or using real-time maps such as Google maps. VSN can help optimize this by establishing social networks between the fleet of vehicles and each of the vehicles with nearby roads. This enables a dynamic monitoring of road conditions in real-time where vehicles can share the road condition and update their routes accordingly. This can help optimize vehicle routing to reduce time and cost. \\ 

GNN can also be an effective tool to find solutions to VRP-related problems by directly generating graph representation of the established social networks, rather than using traditional optimization techniques such as branch and bound, metaheuristics, or exact algorithms, which can be computationally expensive. Each location can be represented as a node in the graph, and the edges between nodes represent the possible routes between locations. In addition, the problem's constraints, such as vehicle capacities and time windows, can be encoded by incorporating them into the graph's structure or node/edge attributes. Capturing the spatial and temporal relationships between locations can also support the generation of more accurate solutions by incorporating information such as distance, traffic, and weather into the graph's structure or node attributes. For example, the authors of~\cite{nammouchi_generative_2020} proposed a new method to solve the Traveling Salesman Problem (TSP) using a generative graph using graph sampling approach. On the other hand, the paper~\cite{barbecho_bautista_evaluation_2020} addresses the problem of enabling an effective re-routing mechanism once the vehicle joins a VSN. Due to its limited connection time, there is a need for GNN models that can dynamically evaluate and produce optimal routing directions for the vehicle to ensure safe driving. A recent work introduces a deep reinforcement learning framework that employs GAT to optimize routes of electric vehicles considering their battery constraints and charging requirements~\cite {DRL_E_vehicle}. It models customers, depots, and charging stations as nodes and connections between these locations as edges weighted by distance or traveling time.
It is worth noting that all presented studies are recent, and the field is still growing, which opens the door for more promising solutions to be designed.


\subsection{Traffic Data Analysis (TDA)}
This task aims to understand the patterns and trends, gain insight into traffic conditions over time, and hopefully identify bottlenecks. It is vital in supporting transportation planning, infrastructure development, and traffic management decisions in ITS in general and VSN-related tasks. The possible application scenarios that can be experimented with GNN are wide in this area. In addition, when using GNNs for traffic data analysis, several factors can be considered important in order to improve the performance and accuracy of the model (e.g., Graph structure, Spatial/Temporal data, Node/edge features, Attention mechanism, Multi-task learning). An example is TAP~\cite{liu_tap_2022}, a novel proposed approach for analyzing traffic accident data that may be useful for transportation planning and traffic management. It employs GNNs and a multi-task learning approach to improve the understanding of traffic accidents over time and identify patterns in the data. It also shows the ability to predict the likelihood of an accident at a specific location and time and identify the factors that might contribute to the accidents. 

A promising and increasingly demanding goal of traffic data analysis solutions is establishing a sustainable VSN. Achieving that requires addressing a number of factors, including security, privacy, trust, scalability, and many more. The dynamic nature of VSN and its short lifetime make this goal a great challenge. In fact, this is still a growing and promising field. There are few works proposed in this area. Recent work proposes a federated learning GNN model for traffic state estimation (TSE) to better analyze traffic and understand patterns~\cite{ShieldTSE}. The modeled Roadside Units (RSU) as nodes that collect traffic data. The edges represent the spatial relations between these RSUs. The framework handles GNN model training in a distributed manner, RSUs learns traffic data locally, and exchanges the data in real-time through a cloud server, which then trains the GNN model to preserve the privacy of the data by sharing immediate activation rather than raw data. Another related work addresses the fake news problem shared in VSN~\cite{guo_mixed_2022}. It proposes a mixed GNN-based approach for fake news detection in VSN that combines both node-level and graph-level information. The model extracts useful information from the graph structure of the data and contents of the exchanged messages, which improves the performance of the model in detecting fake news communicated across vehicles. 

The field is continuously expanding, driven by a growing demand for solutions that enhance traffic analysis, ultimately improving traffic management and supporting better decision-making.

\begin{table*}[t]
\caption{\textcolor{black}{Representative GNN-based VSN models and qualitative practical characteristics across tasks}}
\label{tab:vsn_gnn_takeaways}
\centering
\scriptsize
\setlength{\tabcolsep}{2pt}

{\color{black}
\begin{tabular}{|>{\centering\arraybackslash}m{2.2cm}|
                >{\centering\arraybackslash}m{1.2cm}|
                >{\centering\arraybackslash}m{2.1cm}|
                m{9.5cm}|}
\hline
\textbf{Tasks} & \textbf{Ref.} & \textbf{Proposed Model} & \textbf{Key Characteristics / Practical Notes} \\ \hline

Flow Prediction 
 &~\cite{chen_aargnn_2022} & AARGNN 
 & \begin{itemize}\itemsep1pt
    \item Scales to city-scale recurrent traffic graphs.
    \item Leverages rich node/edge features and exogenous context.
    \item Offers interpretability via attention on influential segments.
    \item Supports near real-time operation after offline training.
   \end{itemize} \\ \hline

Traffic Forecasting 
 &~\cite{bai_adaptive_2022} & AGCRN 
 & \begin{itemize}\itemsep1pt
    \item Scales to large sensor networks via adaptive adjacency learning.
    \item Relies on long historical series for each sensor.
    \item Learns interpretable latent graphs that reveal correlations.
    \item Enables online multi-step forecasting with modest latency.
   \end{itemize} \\ \hline

Traffic Signal Control 
 &~\cite{Zhong_ProbGNN_2021} & TSC-GNN 
 & \begin{itemize}\itemsep1pt
    \item Scales from isolated intersections to network-wide control.
    \item Requires lane-level traffic states and timing data.
    \item Provides interpretable probabilistic outputs and graph weights.
    \item Produces signal plans at second-level latency for real-time use.
   \end{itemize} \\ \hline

Driving Assistance / Autonomous Vehicles 
 &~\cite{ha_road_2021} & Road-GNN
 & \begin{itemize}\itemsep1pt
    \item Scales to dense urban junctions
    \item Uses road network graph and basic kinematic states of vehicles.
    \item Road segments are easy to visualize and relate to driving maneuvers.
    \item Supports real-time planning with GPU acceleration.
   \end{itemize} \\ \hline

Pedestrian Recognition / Obstacle Detection 
 &~\cite{zhou_ast-gnn_2021} & AST-GNN 
 & \begin{itemize}\itemsep1pt
    \item Scales to dense pedestrian urban scenarios.
    \item Requires high-frequency trajectories and interaction data.
    \item Provides interpretability via spatial–temporal attention.
    \item Enables fast short-horizon predictions for on-board systems.
   \end{itemize} \\ \hline

Trajectory Prediction 
 &~\cite{li_hierarchical_2022} & Hierarchical GNN framework 
 & \begin{itemize}\itemsep1pt
    \item Scales using hierarchical global–local graph modeling.
    \item Requires heterogeneous trajectories and map information.
    \item Interpretable: disentangles global and local contributions.
    \item Achieves latency compatible with near real-time use.
   \end{itemize} \\ \hline

Vehicles Routing Problems (VRP) 
 &~\cite{nammouchi_generative_2020} & GLN-TSP 
 & \begin{itemize}\itemsep1pt
    \item Scales to medium-sized routing graphs via embeddings.
    \item Uses full knowledge of topology and edge costs.
    \item Provides partial interpretability via learned representations.
    \item Suited for offline optimization; limited real-time re-routing.
   \end{itemize} \\ \hline

Traffic Data Analysis 
 &~\cite{liu_tap_2022} & TAP 
 & \begin{itemize}\itemsep1pt
    \item Scales to large accident/traffic graphs in multi-task settings.
    \item Requires historical incident and contextual data.
    \item Interpretable via hotspot and pattern profiling.
    \item Primarily for offline forensic analysis.
   \end{itemize} \\ \hline

Connectivity \& Computing 
 &~\cite{10637671} & Transedge 
 & \begin{itemize}\itemsep1pt
    \item Scales to large IoV edge–cloud systems.
    \item Depends on detailed channel and resource information.
    \item Less interpretable due to deep RL-based offloading policy.
    \item Optimized for low-latency, real-time task offloading.
   \end{itemize} \\ \hline

\end{tabular}}
\end{table*}

\subsection{Connectivity and Computing (C\&C)}
 GNNs can play a pivotal role in addressing challenges associated with the connectivity and computational resource allocation of road users. One of the critical issues in VSN is the frequent handover during vehicle mobility in the IoV, which can lead to handover failure and degraded network performance. Traditional methods for network selection and handover decision-making often fall short in dynamically changing environments. GNN has been proposed as a technical alternative to model these systems and mitigate these challenges. Kumar et al.~\cite{pramod2023reinforcement} proposed a novel fuzzy-based GNN approach that integrates fuzzy logic with a hierarchical graph structure. Their method enhances network selection accuracy by enabling vehicles to assess and choose the most suitable network in real-time, thereby minimizing handover failures and delays and significantly improving the quality of service within IoV communications. Moreover, the Cognitive IoV (CIoV) benefits from GNNs through enhanced software management and escalation capabilities. Wang et al.~\cite{9690605} presented a deep learning-based Software Escalation Prediction (SEP) method leveraging GNNs to model software state logs dynamically, hence predicting necessary upgrades and ensuring system reliability and stability. This approach enhances the cognitive and coordination abilities of the CIoV, facilitating timely software updates critical for intelligent transportation systems. In the context of in-vehicle networks, Time-Sensitive Networking (TSN) can provide deterministic low-latency communication, which is crucial for high-level autonomous vehicles. Sun et al.~\cite{10509567} utilized a GNN-based Deep Reinforcement Learning (DRL) approach for TSN scheduling. By framing the scheduling problem as a delay optimization task, the proposed GAT effectively extracts critical information for improved scheduling accuracy, achieving high-precision offline scheduling with low delays.

Additionally, GNNs play a vital role in resource allocation within Cellular Vehicle-to-Everything (C-V2X) communications. By constructing dynamic graphs representing communication links~\cite{GNN_joint_comm}, GNNs enable vehicles to make informed, localized resource allocation decisions while preserving global feature learning capabilities. This method not only ensures high success rates for Vehicle-to-Vehicle communications but also minimizes interference in Vehicle-to-Infrastructure links, optimizing overall network performance. Graham et al.~\cite{10.1007/978-3-031-47126-1_4} highlight the integration of GNNs with clustering methodologies to address persistent challenges such as sporadic connectivity and transmission delays. By learning hidden spatial and functional patterns, GNNs facilitate a more intelligent organization of network nodes, ultimately enhancing resource allocation in highly dynamic environments. 

Finally, as edge computing continues to improve the performance of transportation systems, GNNs have been utilized in task offloading schemes to minimize response latency in edge computing-enabled transportation systems~\cite{10637671}. An adaptive node placement algorithm, combined with a GNN-enhanced scheme, effectively captures spatial features and optimizes task offloading decisions.

\subsection{Takeaways}
 \vspace{0.2cm}

{\color{black}
This section synthesizes the surveyed studies in the previous section and discusses the messages and trends identified through them. To support our analysis based on the categorization presented in Table~\ref{tab:table2_v2}, Table~\ref{tab:vsn_gnn_takeaways} highlights one representative GNN model per VSN-related task, providing a qualitative summary of its scalability, data requirements, interpretability, and real-time suitability. To sum up, the literature review reveals encouraging trends in the application of GNNs to VSNs.  The following key takeaways summarize the major positive outcomes observed from recent studies:\\
$\bullet$ \textbf{Growing interest in integrating GNNs with VSN frameworks:} Compared to earlier ITS surveys, there is a notable rise in studies that explicitly model VSNs or their subdomains (e.g., IoV, VANETs, or SIoV). Recent research on traffic flow prediction~\cite{chen_aargnn_2022,zhang_automatic_2022} and vehicle routing problems~\cite{nammouchi_generative_2020} demonstrates how graph-based representations effectively capture the communication and relational structures of VSNs. This marks an important shift from theoretical exploration to practical modeling of social vehicular interactions using graph learning.\\
$\bullet$ \textbf{Task-specific suitability of GNN variants:} Different GNN architectures demonstrate complementary strengths across VSN-related tasks. GCNNs have proven to be most effective for spatial reasoning in applications such as traffic forecasting~\cite{bui_spatial-temporal_2022} and anomaly detection~\cite{zhang_automatic_2022}, whereas GRNN and GGNN variants excel in sequential or temporal-dependent tasks such as trajectory prediction and spatio-temporal traffic prediction~\cite{mo_graph_2021, Zhiyong_GGNN_traffic_2020}. Attention-based architectures (e.g., GAT and Transformer hybrids) have demonstrated superior adaptability in applications that require contextual awareness, such as traffic signal control~\cite{Zhong_ProbGNN_2021, Li_DeepImitation_2020} and autonomous driving~\cite{multi_modal_trans}. This confirms that GNN selection is becoming increasingly task-driven and fine-tuned to the characteristics of VSN data.\\
$\bullet$ \textbf{Widespread adoption of hybrid architectures:} A clear trend in the reviewed works is the increasing use of hybrid GNN models that combine the strengths of multiple architectures. For example, studies such as STFGNN~\cite{li_spatial-temporal_2021}, AGCRN~\cite{bai_adaptive_2020}, and TSC-GNN~\cite{Li_DeepImitation_2020} integrate GCNN, GRNN, and attention mechanisms to jointly capture spatial and temporal dependencies. These hybrid approaches consistently outperform single-architecture models by enabling dynamic representation of vehicular interactions and achieving higher robustness in varying traffic conditions.\\
$\bullet$ \textbf{Consistent outperformance of GNN solutions:} Across nearly all application domains reviewed, GNN-based approaches outperform traditional machine learning and non-graph deep learning baselines. For instance, the GDN model~\cite{zhang_automatic_2022} achieved superior accuracy in traffic anomaly detection compared to CNN and RNN counterparts, while the Hierarchical GNN~\cite{li_hierarchical_2022} exceeded other sequence models in trajectory prediction (Fig.~\ref{fig:trajec_pred_result}). This recurring result highlights the effectiveness of graph representation learning for capturing complex interdependencies in dynamic vehicular environments.\\
$\bullet$ \textbf{Enhanced analytical and visual interpretability:} The adoption of attention mechanisms and graph visualization techniques in several models (e.g., ST-MGAT~\cite{tian_st-mgat_2020}, AST-GNN~\cite{zhou_ast-gnn_2021}) has improved the interpretability of learned features and decision reasoning. These mechanisms enable better visualization of node and edge importance, supporting explainability in applications such as driving assistance and pedestrian trajectory prediction. Such interpretability is a key advantage of GNN-based frameworks in safety-critical ITS contexts.\\
$\bullet$ \textbf{Increasing awareness of practical deployment aspects:} Beyond these models' accuracy, a large fraction of the works summarized in Table~\ref{tab:table2_v2} explicitly discuss how their GNN architectures scale to larger road networks or fleets, what types and densities of data they rely on (e.g., dense loop-detector grids, multi-modal perception), and whether the proposed models can be executed under real-time or near real-time constraints. These tendencies appear across multiple task categories, including flow prediction and forecasting (e.g., STFGNN, FASTGNN, AGCRN), traffic signal control (e.g., TSC-GNN, STMARL), trajectory prediction, data analysis (e.g., TAP, Mixed-GNN), and connectivity/computing (e.g., MDP-GAT-DRL, Transedge). Table~\ref{tab:vsn_gnn_takeaways} summarizes these aspects for one representative model per task, but similar trade-offs between scalability, data requirements, interpretability, and real-time suitability are visible throughout the broader set of approaches, indicating a growing awareness of practical deployment requirements in GNN-based VSN research.

These takeaways confirm the steady progress and maturity of GNN research in VSNs. The reviewed studies demonstrate that GNN-based methods are no longer limited to proof-of-concept evaluations but are becoming integral components of advanced ITS frameworks. Nevertheless, none of the existing works has yet modeled a complete, standalone VSN encompassing all of its functional components (vehicles, infrastructure, users, and social relations). Current studies remain largely task-specific or focused on sub-VSN graphs, such as traffic flow, routing, or communication sub-networks. This gap highlights an opportunity for future research direction to develop a unified GNN-driven framework that holistically represents and learns from a fully scaled VSN environment.
}

\section{Open Datasets \& Source Codes}
\label{sec5}
\begin{table*}[]
\centering
\caption{List of datasets on traffic flow, speed data, and vehicle trajectories}
\renewcommand{\arraystretch}{1.3}  
\scriptsize  

\begin{tabular}{|p{1.5cm}|p{1.5cm}|p{2.5cm}|p{2cm}|p{2.3cm}|p{1.5cm}|p{1cm}|p{2cm}|}

\hline

\textbf{Dataset} &
\textbf{Type of Data} &
\textbf{Source} &
\textbf{Paper} &
\textcolor{black}{\textbf{Volume}}&
\textcolor{black}{\textbf{Size}} &
\textcolor{black}{\textbf{Labels}} &
\textcolor{black}{\textbf{Applications}}
  \\ \hline
\multicolumn{8}{|c|}{\textbf{Traffic Flow and Speed Data}} \\ \hline
METER-LA 
&\textcolor{black}{Traffic Speed Data (Loop detectors)} & \url{https://bit.ly/3ExJa8L} 
&\cite{chen_aargnn_2022,tian_st-mgat_2020,ruiz_gated_2020,bui_spatial-temporal_2022}
&\textcolor{black}{over 6 million traffic points}
&\textcolor{black}{$\approx 54~\text{MB}$}
&\textcolor{black}{Yes}
&\textcolor{black}{TFP, TF, TSC, TDA, VRP, C\&C}\\
  \cline{1-8}
PEMS-BAY
&\textcolor{black}{Traffic Speed Data (Loop detectors)}
&\url{https://bit.ly/3ExJa8L} 
&\cite{chen_aargnn_2022,tian_st-mgat_2020} 
&\textcolor{black}{around 17 million data points}
&\textcolor{black}{$\approx 130~\text{MB}$}
&\textcolor{black}{Yes}
&\textcolor{black}{TFP, TF, TSC, TDA, VRP, C\&C}\\
\cline{1-8}
PeMS08 
&\textcolor{black}{Traffic Flow Data (Loop detectors)} 
& \url{https://bit.ly/4hr5uPQ} &~\cite{PeMS03_04_07_08,guo_attention_2019,zhang_fastgnn_2021}
&\textcolor{black}{PeMS08: around 18,000 datapoints}
&\textcolor{black}{$\approx 117.5~\text{MB}$}
&\textcolor{black}{Yes}
&\textcolor{black}{TFP, TF, TSC, TDA}
\\ 
\cline{1-8}
PeMSD8
& Traffic Flow Data (Loop detectors)
&\url{https://bit.ly/3W2z9nb} 
&~\cite{huang2020lsgcn, pmlr_v139_pal21b, ma2024spatio} 
&\textcolor{black}{17,856 data points}
&\textcolor{black}{$\approx 55~\text{MB} $}
&\textcolor{black}{Yes}
&\textcolor{black}{TFP, TF, TSC, TDA, VRP}
\\ \cline{1-8}
Seattle Loop data 
&\textcolor{black}{Traffic speed (Loop detectors)}
&\url{https://bit.ly/3GkgwVS}
&~\cite{cui2019traffic,cui2018deep,jiang_BigData_2022} 
&\textcolor{black}{$\approx 34$ million samples}
&\textcolor{black}{$\approx 273.9~\text{MB} $}
&\textcolor{black}{Yes}
&\textcolor{black}{TFP, TF, TSC, TDA, C\&C}
\\ 
\cline{1-8}
Shanghai Traffic Speed (SHSpeed)
&\textcolor{black}{Traffic speed data (GPS)} 
&\url{https://bit.ly/3GGu4w2} 
&~\cite{wang2018efficient, jiang_BigData_2022} 
&\textcolor{black}{Full dataset (private): $\approx3.51$ billion raw GPS points, testbed (public): points from selected 156 road segments}
&\textcolor{black}{Full dataset (private): $310$ GB, testbed (public): $\approx6.8$MB}
&\textcolor{black}{Yes}
&\textcolor{black}{TFP, TF, VRP, TDA, C\&C}
\\
\cline{1-8}

\end{tabular}%

\begin{tabular}{|p{1.5cm}|p{1.5cm}|p{2.5cm}|p{2cm}|p{2.3cm}|p{1.5cm}|p{1cm}|p{2cm}|}

\hline


\multicolumn{8}{|c|}{\textbf{Vehicle Trajectories}} \\ \hline
V2X-Seq
&\textcolor{black}{Vehicles Trajectories (LiDAR, Camera)}
&\url{https://bit.ly/4di5PS2}
&~\cite{V2X-Seq}
&\textcolor{black}{Perception: 95 scenarios, Forecasting: 210,000 scenarios}
&\textcolor{black}{}
&\textcolor{black}{Yes}
&\textcolor{black}{TFP, TF, TSC, DA/AV, TP, VRP, TDA, C\&C}
\\ \cline{1-8}
Carolinas Highway Dataset(CHD)
&\textcolor{black}{Vehicles Trajectories (standard camera)}
&\url{https://bit.ly/4di4cDU}
&~\cite{CHD} 
&\textcolor{black}{22 Videos, 1.6 million frames}
&\textcolor{black}{$\approx 1.71~\text{GB} $}
&\textcolor{black}{Yes}
&\textcolor{black}{TFP, TF, TSC, DA/AV, TP, VRP, TDA, C\&C}
\\ 
\cline{1-8}
CitySim
&\textcolor{black}{Vehicles Trajectories (Drone)}
&\url{https://bit.ly/3TvrT4M} 
&~\cite{citysim} 
&\textcolor{black}{$1,140$ min Video records,$\approx 2$ million frames}
&\textcolor{black}{} 
&\textcolor{black}{Yes}
&\textcolor{black}{TFP, TF, TSC, DA/AV, TP, VRP, TDA, C\&C}
\\ 
\cline{1-8}
T-Drive
&\textcolor{black}{Taxi trajectories (GPS)}
&\textcolor{black}{a small sample of the dataset: }\url{https://bit.ly/4jNmVvL}
&~\cite{Yuan_T_Drive_2010,Yuan_Driving_2011}
&\textcolor{black}{$\approx 790$ million GPS points, the small sample is $15$ million data points}
&\textcolor{black}{the small sample $\approx 146$ MB}
&\textcolor{black}{Yes}
&\textcolor{black}{TFP, TF, TP, VRP, TDA}
\\
\cline{1-8}
TaxiBJ
&\textcolor{black}{Inflow/outflow data of taxicab(GPS)} 
&\url{https://bit.ly/4gGU5tY} 
&~\cite{mourad2019astir,jiang_BigData_2022}
&\textcolor{black}{$\approx 25.87$ million inflow/outflow data points}
&\textcolor{black}{$\approx71.08$ MB}
&\textcolor{black}{Yes}
&\textcolor{black}{TFP, TF, VRP, TDA, C\&C}
\\
\cline{1-8}


OpenStreetMap 
&Road network data (GPS, Camera, Satellite)
&\url{https://bit.ly/3QnFqJx} 
&~\cite{chen_robust_2021}
&\textcolor{black}{Nodes: $\approx 7 \text{ billion}$ , Ways: $\approx 957 \text{ million}$, Relations:$\approx 24.6 \text{ million}$}
&\textcolor{black}{$\approx 2 \text{TB}$}
&\textcolor{black}{Yes}
&\textcolor{black}{TFP, TF, VRP, TSC, TDA, C\&C}
\\
\cline{1-8}

DiDi Chuxing
&Trajectories data (GPS)
&\url{https://bit.ly/4i7fs9t} 
&~\cite{jiang_BigData_2022,chen_robust_2021}
&\textcolor{black}{$\approx 7$ billion ride request records, $\approx 1$ billion GPS trajectory points} 
&\textcolor{black}{}
&\textcolor{black}{No}
&\textcolor{black}{TFP, TF, TP, VRP, TDA, C\&C}
\\
\cline{1-8}

BLVD 
&High-resolution traffic video clips (Camera + LiDAR)
&\url{https://bit.ly/3jUGEPm} 
&~\cite{li_hierarchical_2022}
&\textcolor{black}{654 Video Clips and $\approx 120,000$ frames}
&\textcolor{black}{$42.7$GB}
&\textcolor{black}{Yes}
&\textcolor{black}{All}
\\
\cline{1-8}

Road-Intersec-Traffic 
& Road Intersection traffic(LiDAR)
&\url{https://dx.doi.org/10.21227/vm2m-ar04} 
&~\cite{Annotated_3DLiDAR_2025},~\cite{AIDriven_traffic_2025}
&\textcolor{black}{$102$ scenarios, $\approx 2,591$ frames}
&\textcolor{black}{$\approx 42$ GB }
&\textcolor{black}{Yes}
&\textcolor{black}{All}
\\
\cline{1-8}

 \end{tabular}%

\label{tab:my-table1}
\end{table*}



\begin{table*}[]
\centering
\caption{List of social network datasets}
\renewcommand{\arraystretch}{1.3}  
\scriptsize  

\begin{tabular}{|p{1.5cm}|p{1.5cm}|p{2.5cm}|p{2cm}|p{2.3cm}|p{1.5cm}|p{1cm}|p{2cm}|}

\hline

\textbf{Dataset} &
\textbf{Type of Data} &
\textbf{Source} &
\textbf{Paper} &
\textcolor{black}{\textbf{Volume}}&
\textcolor{black}{\textbf{Size}} &
\textcolor{black}{\textbf{Labels}} &
\textcolor{black}{\textbf{Applications}}
  \\ \hline




Stanford Dataset Collection 
&List of Social Networks 
&\url{http://snap.stanford.edu/data/} 
&~\cite{leskovec_learning_nodate}
&\textcolor{black}{over $50$ distinct network datasets}
&\textcolor{black}{varies across datasets}
&\textcolor{black}{Yes}
&\textcolor{black}{All}
\\ 
\cline{1-8}




Refined ego-Facebook dataset 
&Crowdsourcing 
&\url{https://snap.stanford.edu/data/egonets-Facebook.html} 
&~\cite{hamrouni_low-complexity_2022}
&\textcolor{black}{$\approx 4,000$ nodes, $\approx 88,000$ edges}
&\textcolor{black}{$950.7$ KB}
&\textcolor{black}{Yes}
&\textcolor{black}{All}
\\ 
\cline{1-8}

Amazon 
&Social network 
&\url{http://jmcauley.ucsd.edu/data/amazon/} 
&~\cite{cen_representation_2019} 
&\textcolor{black}{$\approx 233.1$ million reviews}
&\textcolor{black}{$34$ GB}
&\textcolor{black}{Yes}
&\textcolor{black}{All}
\\
\cline{1-8}

YouTube 
&Social network 
&\url{https://stanford.io/41lmcdH} 
&~\cite{houyoutube} 
&\textcolor{black}{$\approx 1.1$ million nodes, $\approx 3$ million edges}
&\textcolor{black}{$11.13$ MB}
&\textcolor{black}{Yes}
&\textcolor{black}{All}
\\ 
\cline{1-8}

Higgs Twitter 
&Social network 
&\url{https://stanford.io/3ZfRP5F} 
&~\cite{cen_representation_2019} 
&\textcolor{black}{$\approx 456,626$ nodes, $\approx 15$ million edges}
&\textcolor{black}{$61.86$ MB}
&\textcolor{black}{No}
&\textcolor{black}{All}
\\ 
\cline{1-8}


\end{tabular}%
\label{tab:my-table2}
\vspace{-2mm}
\end{table*}









\begin{table*}[]
\centering
\caption{Other datasets suitable for modeling VSN-related Tasks with graph-based approaches}
\renewcommand{\arraystretch}{1.3}  
\scriptsize  

\begin{tabular}{|p{1.5cm}|p{1.5cm}|p{2.5cm}|p{2cm}|p{2.3cm}|p{1.5cm}|p{1cm}|p{2cm}|}

\hline

\textbf{Dataset} &
\textbf{Type of Data} &
\textbf{Source} &
\textbf{Paper} &
\textcolor{black}{\textbf{Volume}}&
\textcolor{black}{\textbf{Size}} &
\textcolor{black}{\textbf{Labels}} &
\textcolor{black}{\textbf{Applications}}
  \\ \hline

\multicolumn{8}{|c|}{\textbf{Accidents Data}} \\ \hline
ML4RoadSafety 
&Accidents analysis 
&\url{https://bit.ly/3BayQ4Q} 
&~\cite{nippani_graph_2023}
&\textcolor{black}{$\approx9$ million incidents}
&\textcolor{black}{$6.7$ GB}
&\textcolor{black}{Yes}
&\textcolor{black}{TSC, DA/AV, VRP,TDA}
\\
\cline{1-8}

US Accidents (2016 - 2021)
&Car accidents data 
&\url{https://bit.ly/3CnAuxy} 
&~\cite{jiang_BigData_2022} 
&\textcolor{black}{$\approx2.8$ million accident reports}
&\textcolor{black}{$\approx1.2$ GB}
&\textcolor{black}{Yes}
&\textcolor{black}{TFP, TF, TSC, DA/AV, VRP, TDA}
\\
\cline{1-8}

\end{tabular}%

\begin{tabular}{|p{1.5cm}|p{1.5cm}|p{2.5cm}|p{2cm}|p{2.3cm}|p{1.5cm}|p{1cm}|p{2cm}|}

\hline


\multicolumn{8}{|c|}{\textbf{Bike Data}} \\ \hline
BikeNYC
&Bike data (GPS)
&\url{https://bit.ly/4gJfF12} 
&~\cite{jiang_BigData_2022},~\cite{mourad2019astir} 
&\textcolor{black}{Over $2$ million trip records (latest)} 
&\textcolor{black}{$\approx 414.2$ MB}
&\textcolor{black}{Yes}
&\textcolor{black}{TFP, TF,TP, VRP, TDA}
\\
\cline{1-8}
BikeDC 
&Bike  data (Docking Station Logs)
&\url{https://bit.ly/3Qgp3xu} 
&~\cite{jiang_BigData_2022} 
&\textcolor{black}{$285732$ trip records (latest)}
&\textcolor{black}{$\approx52.8$ MB}
&\textcolor{black}{Yes}
&\textcolor{black}{TFP, TF, TP, VRP, TDA}
\\ 
\cline{1-8}

BikeChicago 
&Bike trip data (Docking Station Logs)
&\url{https://bit.ly/3D2hpVp} 
&~\cite{jiang_BigData_2022} 
&\textcolor{black}{$138690$ trip records (latest)}
&\textcolor{black}{$\approx28.6$ MB}
&\textcolor{black}{Yes}
&\textcolor{black}{TFP, TF, TP, VRP, TDA}
\\
\cline{1-8}

NYC Taxi data 
&Taxi trips records (GPS + onboard trip records) 
&\url{https://bit.ly/4ifCiv0} 
&~\cite{ma_forecasting_2022} 
&\textcolor{black}{over 26 million taxi trip records (latest)}
&\textcolor{black}{over 560 MB }
&\textcolor{black}{Yes}
&\textcolor{black}{TFP, TF, TP, VRP, TDA}
\\
\cline{1-8}

\multicolumn{8}{|c|}{\textbf{Weather Data}} \\ \hline
NOAA 
&Climate data (multiple environmental sensors)
&\url{https://bit.ly/3ClgNq7} 
&~\cite{isufi_graph-time_2021} 
&\textcolor{black}{Not disclosed (updated regularly)}
&\textcolor{black}{A weekly archive is $\approx600$ MB}
&\textcolor{black}{Yes}
&\textcolor{black}{TFP, TF, DA/AV, VRP, TDA}
\\ 
\cline{1-8}

Molene
&Climate data (mult-environmental sensors)
&\url{https://bit.ly/3VLyupJ} 
&~\cite{isufi_graph-time_2021} 
&\textcolor{black}{$\approx744$ hourly observations}
&\textcolor{black}{$\approx1.09$ MB}
&\textcolor{black}{Yes}
&\textcolor{black}{TFP, TF, DA/AV, VRP, TDA }
\\ 
\cline{1-8}
\end{tabular}%
\label{tab:my-table3}
\end{table*}
This section presents a list of the commonly used datasets in the surveyed studies. Each dataset is illustrated with the type of data it presents, the public source, if applicable, and the main reference. We have gathered an extensive list of datasets related to the topic of interest and provided a great summary to researchers and practitioners in finding suitable datasets for their experiments or getting a rough review of the existing datasets for applications under the ITS field in general. 
It is worth mentioning that there is a lack of rich public datasets that are collected from VSNs. Most of the data collected in the literature are not published due to users' data privacy; examples are~\cite{li2020drive2friends}. However, since this paper focuses on applying GNN to VSN, we focused on datasets that are public, can be modeled as graphs, and provide useful features for VSN. Example work is~\cite{9446513}, which proposes a GNN-based approach to model the social relationships from device features as well as social relations in order to extract and understand such complex relations.

A complete list of the available datasets is presented in Tables~\ref{tab:my-table1},{\color{black} ~\ref{tab:my-table2}, and ~\ref{tab:my-table3}}. They are categorized into six categories:
\paragraph{{\color{black}Traffic Flow and Speed Datasets}} are collected from distributed road sensors in different locations over a specific time range. {\color{black}They have mainly been  used for traffic flow prediction and traffic forecasting tasks}. For example, the famous METER-LA and PEMS datasets have been used for predicting traffic volume and vehicle speed forecasting~\cite{zhang_fastgnn_2021, tian_st-mgat_2020, zhou_variational_2021}. {\color{black} Other potential applications are also possible, as mentioned in Table~\ref{tab:my-table1}}
\paragraph{Vehicles' Trajectory Datasets} represent data that is available mainly for trajectory analysis and prediction studies. {\color{black} As shown in Table~\ref{tab:my-table1}, datasets under this category are collected using GPS tracking systems, Drone, Camera, LiDAR, or a combination of multiple of these sensors. Sensors can be available in vehicles themselves, in road infrastructure, or even on smartphones.} For example, the authors of~\cite{chen_robust_2021} used the road network and trajectories of two cities from OpenStreetMap and DiDi Chuxing for the purpose of evaluating the ability of their proposed framework to capture the needed characteristics of the road in order to make better predictions. {\color{black} In the recent work~\cite{Annotated_3DLiDAR_2025,AIDriven_traffic_2025}, the dataset was collected using a simulated lidar sensor installed in road infrastructure to observe the traffic in road intersections. Potential applications can be wide as demonstrated in Table~\ref{tab:my-table1}.} 
\paragraph{Social Network Datasets}{\color{black}These data is widely popular in data science and deep learning. In the context of this paper, although social network data is not directly related to VSN, synthetic VSN data can be generated to mimic real-world human to infrastructure or vehicles interaction as well the social interaction through these networks represented and processed as graphs. As shown in Table~\ref{tab:my-table2}, VSN-related applications using these datasets are wide, covering all applications presented in this survey. For example, authors of~\cite{houyoutube,cen_representation_2019} utilized Twitter, Amazon, and YouTube social networks to evaluate the performance of the proposed GATNE in representing heterogeneous data networks.} 


\paragraph{Accident datasets} can also be used to analyze and extract useful patterns, such as understanding vehicle relationships prior to accidents. Countrywide car accident data collected from a number of US states is presented in the paper~\cite{jiang_BigData_2022}, which lists different types of traffic datasets based on the type of data collected, along with their references. {\color{black}Refer to Table~\ref{tab:my-table3} for Accident datasets and the below datasets categories}.
\paragraph{Taxi and Bike Datasets} can be collected using different approaches, as presented in~\cite{jiang_BigData_2022}. For example, it can be collected from highway traffic sensors or by collecting data from vehicles over a specified period of time. As with any traffic data, it can be a great data source to evaluate the GNN models. For instance, in~\cite{ma_forecasting_2022}, the authors presented the Deep NYC Taxi Bike dataset for evaluating GNN-based spatiotemporal models to improve traffic prediction tasks using a multi-model strategy. The dataset integrates region-based traffic data collected from public transportation in New York City (NYC) and other demanding metrics between 2019 and 2020.

\paragraph{Weather Datasets} can be a great resource for analyzing and detecting climate change-related tasks using GNN. Although the papers reviewed in this area can go beyond the scope of this paper, the dataset used and the proposed models could be useful for future GNN for VSN applications. For instance, the authors of~\cite{isufi_graph-time_2021} have used NOAA and Melone datasets to evaluate the GTCNN model architecture they proposed. NOAA contained climate temperature data, which was used in the experiment's forecasting task. The data in the Melone dataset was utilized to evaluate the ability of the proposed model to perform precursor-based earthquake detection and classification.


\section{Future Research Directions}
\label{sec6}
GNNs have achieved remarkable success in modeling complex smart mobility and intelligent transportation systems. However, this significant achievement still confronts multiple challenges that prevent guaranteed optimality, scalability, and real-time operation. GNNs still suffer from generalization issues and stability when working with various datasets. In this section, we summarize the main challenges and future directions that researchers may consider in future GNN for smart transportation and VSN applications.

\paragraph{Sophisticated GNN Architectures} The state-of-the-art GNN architectures that have been proposed so far have shown impressive performances in various tasks such as node/graph classification and link prediction. However, in the VSN environment, current architectures face the challenges of implementing large-scale neural networks for delay-intolerant applications and increasing the complexity/size of data over time. Since VSN environments involve multiple possible tasks and communication channels, one possible approach is to integrate multiple GNN architectures that fit well with such dynamic environments in order to achieve better robustness and scalability. GATs are an example of a sophisticated GNN architecture that allows for the efficient processing of large graphs. AARGNN model~\cite{chen_aargnn_2022} employs this architecture for flow prediction tasks, which show promising results. Another example is Graph Transformer Networks (GTNs), which are built on top of Transformers architecture. This sophisticated architecture allows us to learn complex relationships between nodes in graphs, which leads to more robust and accurate predictions. It shows a promising tool for analyzing and understanding the complex data structure in VSN environments. For instance, the authors of STAR~\cite{Yu_STGTN_2020} designed a graph transformer model that presents a promising performance in predicting pedestrian trajectory. Those examples illustrate an encouraging direction of research in designing GNN networks that can handle heterogeneous data and be applied in real time. However, more sophisticated GNN architectures are required to cope with the lack of model architecture explainability. This can greatly benefit the development of robust, sophisticated GNN models that are scalable, dynamic, and explainable enough.
    
\paragraph{Hybrid Learning Approaches} One of the promising directions to address the ITS and smart mobility challenges is to design learning models that combine GNN approaches with other learning models so as to leverage the strengths of the GNN models and algorithms, e.g., to better model the impact of environmental and human factors on traffic patterns. A promising illustration is to combine reinforcement learning with GNN models, which can enhance the autonomy of transportation systems such as smart traffic, autonomous vehicles, and unmanned aerial vehicles. Another example is to combine federated learning with GNN (e.g., FASTGNN~\cite{zhang_fastgnn_2021}, which shows promising performance in mobile edge computing by providing rapid processing and minimization of latency). Another possibility is the integration of generative models with GNN to enhance the graph modeling efficiency and, therefore, produce more robust models. 

The current efforts show the great potential of generative models in enhancing the robustness of solving complex VSN tasks, such as the proposed model in~\cite{Zhong_STGM_2022}. Possible future directions might be exploring the use of generative models with GNN to understand better the social behavior of traffic participants, such as the likelihood of drivers taking certain routes or making certain maneuvers. Generative models may also be used to simulate different smart mobility scenarios, while GNNs can be used to model the social interactions between different drivers. This will lead to a better understanding of the social behaviors in traffic and develop more effective strategies for managing traffic. Another application is by integrating generative models with GNNs in training autonomous vehicles to make more accurate predictions and decisions in complex environments. Hybrid GNNs can be more robust to variations in the data and more adaptable to different types of transportation problems. However, this requires careful consideration of the trade-offs between different models and algorithms, especially in terms of complexity, as well as the design of appropriate training strategies to effectively combine them, considering the unique characteristics and challenges of the transportation domain.
    
\paragraph{GNNs for Multimodal Transportation Systems} The transportation system is a complex environment that consists of various entities, such as roads, railways, airways, and waterways, which are interconnected and interdependent. With the current proposed GNN models, they can only perform tasks on a single mode of transportation or data collected from one city or region. In fact, there is a lack of GNN models that can deal with multiple ITS systems simultaneously (e.g., managing nationwide public transportation means). Building multimodal GNNs for transportation systems is a promising direction that can effectively capture the interdependencies and interactions among different modes, allowing for better predictions, control, and optimization of the transportation system. This approach can help in providing a better mobility option for people and goods. To the best of the authors' knowledge, there is only one paper that introduces a dataset for multimodal transportation demand~\cite{ma_forecasting_2022}. This direction is still growing and needs more research efforts and industry support by providing multimodal datasets that are aligned with real-world situations.
    
\paragraph{Real-world Dataset Availability} Most of the currently available open-source datasets are captured in a specific time range and provide data from a specific city or region. Although they were captured in real time, they could be old data and might not represent realistic data sources that simulate the current real-world situations. Moreover, there is a lack of data that are represented in the form of graphs, as they can greatly enhance the accuracy of GNN models and generate more realistic results for different situations that can occur in the VSN environment. Currently, researchers are using simulators (e.g., Carla, SUMO, and Aimsun) in order to simulate the model performance and data. However, there is a need for more real-life graph-based datasets to better model and capture the different complex situations in more than one region or city. This encourages the establishment of collaboration between governments, industry, and research communities.
    
\paragraph{Privacy, Security, and Trustworthiness}
Although they have shown great potential in improving the privacy, security, and trustworthiness of VSNs, GNNs can be further enhanced to better handle the privacy and security challenges in these networks. For example, GNNs can be used to detect anomalies and attacks, such as fake messages~\cite{guo_mixed_2022} or malicious nodes. GNNs can also be used to enhance privacy-preserving mechanisms by allowing nodes to exchange information without revealing sensitive data. Moreover, GNNs can be utilized to establish trust in VSNs by incorporating reputation systems that assess the behavior of participating nodes. In a nutshell, the future direction of GNNs for improving the privacy, security, and trustworthiness of VSNs is promising, and it will continue to evolve as new challenges arise.

 \paragraph{Practical Implementation Challenges of GNNs in VSNs} While GNNs hold great potential for solving complex problems in VSNs, their practical implementation faces significant challenges. As highlighted earlier, the lack of large-scale, real-world datasets and the need for more scalable and real-time capable GNN architectures hinder their practicality. Issues such as privacy, security, and trust management remain major concerns when deploying GNNs in the decentralized environment of VSNs. Furthermore, the concept of VSNs itself is still in its beginning and has not yet reached maturity in real-world applications. The widespread deployment of VSNs remains limited, making it difficult to implement and evaluate GNN-based models effectively. Beyond these technical limitations, communication and computation resources in VSN environments pose substantial challenges. Real-time data exchange between vehicles requires highly efficient communication protocols and low-latency communication networks. Ensuring constant connectivity, particularly in dynamic and high-mobility scenarios, is crucial for the seamless operation of GNN models in VSNs. Additionally, processing large-scale, dynamic graph data in real-time may exceed the computational resources available in current vehicular systems. Addressing these challenges will require coordinated efforts across academia, industry, and governmental institutions to bring the vision of GNN-enabled VSNs closer to practical reality.

\section{Conclusion}
\label{sec7}
{\color{black} This paper provides an extensive survey that explores the use of GNNs to enable and advance applications within VSNs. By examining the integration of GNNs within complex and dynamic systems, we have demonstrated their capacity to effectively model the complex interactions and data patterns crucial for ITS and smart mobility. Additionally, we have highlighted GNNs' ability to manage the dynamic and often non-linear relationships characteristic of transportation data, making them exceptionally well-suited for a range of VSN-related applications. Our analysis revealed several key takeaways. Hybrid spatial–temporal GNNs are increasingly favored for their ability to capture both structural and temporal dynamics. Different GNN variants show clear task-specific suitability: attention-based models excel in interaction modeling, graph convolutional networks in flow prediction, and hierarchical architectures in trajectory reasoning. GNN-based solutions also consistently outperform traditional Euclidean approaches by exploiting relational and dynamic graph structures. However, despite these advances, existing studies remain largely task-specific, with no work yet modeling a complete, standalone VSN that integrates all functional components.

Although the deployment of GNNs across various VSN-related tasks has shown promising results, the field remains in an early stage. Several challenges remain unsolved, particularly with respect to achieving scalable architectures capable of supporting large and multimodal networks, integrating robust security and privacy mechanisms, and enabling real-time inference under dynamic mobility conditions. Addressing these challenges requires coordinated efforts across academia, industry, and governmental institutions to realize the full potential of GNN-enabled VSNs and guarantee their practical implementation and effective integration into the future of smart mobility and vehicular networks.}

\bibliography{GNN_VSN_research}
\bibliographystyle{ieeetr}
\balance
\end{document}